\documentclass[sigconf]{acmart} %
\AtBeginDocument{%
  }

\setcopyright{acmlicensed} %
\copyrightyear{2018} %
\acmYear{2018} %
\acmDOI{XXXXXXX.XXXXXXX} %
\acmConference[Conference acronym 'XX]{Make sure to enter the correct
  conference title from your rights confirmation email}{June 03--05,
  2018}{Woodstock, NY}  %
\acmISBN{978-1-4503-XXXX-X/2018/06}  %

\usepackage{algorithm}      %
\usepackage{algpseudocode}    %
\usepackage{graphicx}
\usepackage{textcomp}
\usepackage{multicol}
\usepackage{multirow}
\usepackage{tabularx}
\usepackage{makecell}
\usepackage{xspace}
\usepackage{threeparttable}
\usepackage[normalem]{ulem}
\usepackage{bm} 
\newcommand{\ie}{i.e.,\xspace}
\newcommand{\eg}{e.g.,\xspace}
\newcommand{\obased}{optimization-based\xspace}
\newcommand{\lbased}{learning-based\xspace} %
\usepackage[caption=false,font=footnotesize]{subfig}

\usepackage[utf8]{inputenc} %
\usepackage[T1]{fontenc}    %
\usepackage{hyperref}       %
\usepackage{url}            %
\usepackage{booktabs}       %
\usepackage{nicefrac}       %
\usepackage{microtype}      %
\usepackage{xcolor}         %
\usepackage{comment}
\newcommand{\tool}{HarmonicAttack\xspace}
\newcommand{\newpar}[1]{\noindent\textbf{#1.}\xspace}

\newcommand{\dtrain}{$\mathcal{D}_{\text{train}}$\xspace}
\newcommand{\deval}{$\mathcal{D}_{\text{eval}}$\xspace}
\newcommand{\audiomarkbench}{AudioSquareAttack\xspace}

\begin{document}

\title{\tool: An Adaptive Cross-Domain Audio Watermark Removal
}

\author{Kexin Li}
\authornote{Equal contributions.}
\affiliation{
  \institution{University of Toronto}
  \country{Canada}
}

\author{Xiao Hu}
\authornotemark[1] 
\affiliation{
  \institution{University of Toronto}
  \country{Canada}
}

\author{Ilya Grishchenko}
\affiliation{
  \institution{University of Toronto}
  \country{Canada}
}

\author{David Lie}
\affiliation{
  \institution{University of Toronto}
  \country{Canada}
}

\renewcommand{\shortauthors}{Li et al.}

\begin{abstract} %
The availability of high-quality, AI-generated audio raises security challenges such as misinformation campaigns and voice-cloning fraud. A key defense against the misuse of AI-generated audio is by watermarking it, so that it can be easily distinguished from genuine audio. Those seeking to misuse AI-generated audio may attempt to remove audio watermarks, so studying effective watermark removal techniques is critical to objectively evaluate the robustness of audio watermarks.
Previous watermark removal schemes typically assume access to the target watermark detector during the removal process. This assumption is often impractical, which may lead to a false sense of confidence in current watermark schemes.

We introduce \tool{}, a novel audio watermark removal method that requires no access to the target watermark algorithm. It only needs a number of original and watermarked samples to train a general model capable of removing watermarks from audio samples. We also find that training samples do not need to share the same distribution as target samples, as our attack generalizes to out-of-distribution samples with minimal degradation. \tool{} employs a dual-path convolutional autoencoder operating in temporal and frequency domains, along with GAN-style training, to separate the watermark from the original audio.

Compared with existing watermark removal attacks, \tool{} is more effective at removing watermarks from state-of-the-art schemes, including AudioSeal, WavMark, SilentCipher, and AudioMarkNet, while maintaining high perceptual quality. Although \tool{} is trained on the LibriSpeech dataset against AudioSeal, it generalizes across unseen datasets and watermarking schemes. For instance, on the VCTK dataset, \tool{} achieves a 92\% attack success rate (ASR) against AudioMarkNet, substantially outperforming the best baseline at 38\%. On the FMA dataset, \tool{} reaches 100\% ASR against all watermarks, whereas the best baseline achieves only 2\% against AudioSeal and 44\% against WavMark.

\end{abstract}


\maketitle

\section{Introduction} \label{sec:intro}

Recent advancements in generative audio models~\cite{kimiteamKimiAudioTechnicalReport2025,chuQwen2AudioTechnicalReport2024} have revolutionized sound creation, manipulation, and interaction across tasks such as text-to-speech, music generation, and audio enhancement by enabling realistic and high-quality audio synthesis. This growing realism raises serious concerns regarding misinformation, impersonation, and fraud~\cite{park2024ai}. For instance, voice cloning enabled by AI-generated speech can be exploited for malicious purposes such as fraudulent phone calls to financial institutions, large-scale social engineering attacks, or automated voice scams that convincingly mimic trusted individuals~\cite{UniversityWaterloo2025_WatermarksDefenceDeepfakes, guardian2024wppdeepfake}. Fraud using AI-generated speech has already led to multi-million dollar financial losses~\cite {brewster2021voiceclone35m}.

One effective solution to address these challenges is \emph{watermarking}~\cite{audioseal,chenWavMarkWatermarkingAudio2024,liuDetectingVoiceCloning2023}. Modern watermarks modify the audio generation process such that each synthesized output is automatically encoded with a unique, hidden footprint, which can be extracted by a corresponding watermark detector.

There are several fundamental design principles behind modern watermarking systems. First, detectability requires embedding a unique signal that is unlikely to naturally occur in audio content, ensuring reliable identification of the watermark signal. Second, the watermark should not degrade the perceptual quality of the original audio. To do so, modern audio watermarking systems typically leverage the \emph{psychoacoustic masking} effect~\cite{swansonRobustAudioWatermarking1998, spreadspectral}, embedding watermark signals into time-frequency bins that are less perceptually sensitive due to the presence of stronger original acoustic energy, which renders the watermark perturbations inaudible to human listeners. Third, robustness requires that the embedded signal remain difficult to remove. If watermarks can be erased without significant distortion, the detection guarantee is weakened. Therefore, testing state-of-the-art watermarks against removal is important, as it provides direct insight into how watermarking systems can be improved. 

Guided by the design principles of watermarking, we introduce three key watermark removal objectives: (1) \emph{precise removal:} it must precisely learn and model watermark signals to actively remove the watermark residual, rather than relying on generic distortions; (2) \emph{localization:} it must identify where watermark signals are concentrated, which typically corresponds to psychoacoustically masked time-frequency bins; (3) \emph{quality:} it must preserve the perceptual quality of the watermarked audio to remain practical. 

Existing approaches fail to jointly address these objectives, often trading off removal strength for audio quality or lacking the ability to localize watermark embeddings.
Signal-processing attacks, such as bandpass filtering and lossy compression \cite{oreillyDeepAudioWatermarks2025}, often achieve especially low attack success rates on modern robust watermarks. 
Attack effectiveness is better for the \obased methods (\eg HopSkipJump and Square attacks \cite{andriushchenkoSquareAttackQueryefficient2020,liuAudioMarkBenchBenchmarkingRobustness}), but they lack practicality as they require repeated query access to the watermark detector, and in some settings, confidence scores or hard-label feedback. 
These existing attacks also fail to effectively remove modern watermarks without introducing audible artifacts.

To overcome the limitations of existing removal methods, we introduce an objective-driven \tool, the first \lbased black-box audio watermark removal attack that does not require access to the watermark detector.
The attacker obtains only data samples of the original audio and their corresponding watermarked versions.
This reflects a realistic deployment scenario in which original/watermarked audio pairs are publicly accessible, either through watermarked audio leakage on the public internet or through a public-facing watermark generation API that an adversary can query after registering as a benign user.
Compared to prior methods, \tool achieves higher attack success rates while maintaining high perceptual audio quality. 

Since our method only requires paired samples, we design it to learn watermark characteristics independently of the underlying audio distribution, enabling strong transfer to unseen data and watermarking schemes. This \textit{generalization} capability arises because modern watermarking schemes are constrained by the design principles, specifically the need to preserve perceptual quality and thus follow psychoacoustic masking~\cite{swansonRobustAudioWatermarking1998, spreadspectral}. As a result, watermark signals tend to cluster in similar time-frequency components of the spectrogram, particularly around source-dominant frequency bands. By learning both the watermark structure and these shared embedding locations, our attack transfers effectively across different audio content and watermarking schemes.

\tool employs a dual-path autoencoder architecture that jointly processes waveform and spectrogram representations of the watermarked audio, capturing temporal and spectral characteristics of watermarks. This design enables the model to learn the watermark structure and localize watermark-embedded components in the time-frequency space, addressing both precise removal and localization watermark removal objectives. In addition, we introduce a multi-component loss function that explicitly optimizes the model towards these objectives: a disentanglement loss encourages the model to learn the structure of and remove watermark signals (precise removal objective), a localization loss guides the model to focus on masked time-frequency components where watermark energy is concentrated (localization objective), and a reconstruction loss preserves perceptual fidelity (quality objective). Finally, we incorporate adversarial (GAN-style) training~\cite{goodfellow2014generativeadversarialnetworks, sharma2024generative}, where a discriminator encourages the generator to produce outputs that are indistinguishable from original audio, ensuring that the watermark removal generator does not introduce perceptual artifacts.

Together, these components enable \tool to achieve effective, high-quality, and generalizable watermark removal. Unlike optimization-based attacks whose cost scales with input length, our method performs removal in near real time and remains efficient even for long audio samples.

To summarize, the paper makes the following contributions:
\begin{itemize}
    \item We introduce the first \lbased black-box audio watermark removal system that learns watermark structure and localization directly from the data, without requiring access to the watermarking algorithm or detector.
    \item We design a dual-path architecture and a multi-objective loss function that jointly address the three key objectives of watermark removal: precise removal enabling watermark signal modeling, localization in masked time-frequency bins, and high perceptual quality preservation.
    \item We evaluate \tool across multiple datasets and watermarks, and show strong transferability in cross-dataset and cross-watermark evaluations. In particular, when trained on speech data from LibriSpeech~\cite{librispeech}, \tool's universal watermark removal model achieves 100\% watermark removal against all evaluated watermarking schemes on unseen music data from the Free Music Archive (FMA)~\cite{fma_dataset}, while preserving high audio quality---ViSQOL\footnote{ViSQOL stands for Virtual Speech Quality Objective Listener, which is an objective metric that estimates perceived audio quality on both speech and music using a MOS-LQO score ranging from 1 to 5, where higher values indicate better quality~\cite{chinen2020visqolv3opensource}. } is around 4.33.
    This performance exceeds that of state-of-the-art removal methods.
    For AudioMarkNet~\cite{AudioMarkNet}, the strongest watermark in our experiments, the best baseline removal method achieves only 38\% attack success rate (ASR) on the VCTK dataset~\cite{yamagishi2019vctk}, while \tool's ASR is 92\%.
\end{itemize}

\section{Background} \label{sec:bg}

\subsection{Audio Watermarking Schemes}

Watermarking systems \cite{audioseal, AudioMarkNet} typically consist of an embedder that inserts the watermark and a detector that verifies its presence. 
Thus, when embedded in AI-generated audio, watermarks can help identify their origin.
The fundamental challenge in this domain lies in balancing two competing requirements: \emph{fidelity} (maintaining audio quality so that the watermark is imperceptible) and \emph{robustness} (surviving audio manipulations and performing consistently across different audio types, such as speech and music). Achieving higher robustness often means that the watermark has higher energy, which generally weakens its fidelity. Therefore, modern watermarking schemes aim to equalize both requirements by making subtle modifications to audio signals that take advantage of the limitations in the human auditory system.

Audio watermarking techniques can be categorized based on the domain in which they operate. Time-domain watermarking directly modifies the audio waveform by adding watermark signals. Frequency-domain watermarking transforms the audio signal to reveal its frequency components, typically using the Fast Fourier Transform (FFT) or Short-Time Fourier Transform (STFT)~\cite{STFT}, and modifies the spectral coefficients. Transform-domain watermarking utilizes signal representations such as wavelet transforms, which can provide multi-resolution analysis. Modern watermarking methods are typically hybrid-domain. For instance, AudioSeal \cite{audioseal} uses a generator---detector architecture where the generator produces an additive watermark waveform directly in the time domain. However, its loss function accounts for the time-frequency features of the input, emphasizing the frequency-band signals where the original signal has high energy. This allows the approach to leverage the auditory masking property to improve fidelity. Similarly, AudioMarkNet \cite{AudioMarkNet} exploits psychoacoustic masking by embedding perturbations in low-frequency and high-energy regions, where they remain imperceptible while being tightly coupled with the audio content, thereby improving imperceptibility.

\subsection{Audio Watermark Removal}

Approaches to watermark removal can be categorized based on their technical methodology. Conventional signal processing~\cite{oreillyDeepAudioWatermarks2025} employs operations like signal-level distortions, denoising, codecs, or noise addition in an attempt to eliminate watermark signals without prior knowledge of the watermarking system. Some methods further rely on additive interference~\cite{fakhr2012robust, shih2017digital, hartung1999spread}, introducing perturbations designed to reduce detector reliability while preserving perceptual quality.  However, the effectiveness and generalizability of these approaches are often limited by their reliance on fixed, hand-crafted transformations.
Optimization-based methods such as HopSkipJump and Square attacks, adapted to audio watermark removal in a recent study~\cite{liuAudioMarkBenchBenchmarkingRobustness}, iteratively optimize the noise patterns imposed on the watermarked samples. The optimization process is guided by the watermark detection confidence score, accelerating the convergence. However, these methods are based on the strong assumption that the detector remains continuously accessible to the attacker. 
In practice, such access can be easily restricted or revoked, making the attack unreliable. The iterative querying process also incurs high computational overhead and leads to long attack times.

\begin{figure*}[ht!]
    \centering
    \includegraphics[width=.9\textwidth]{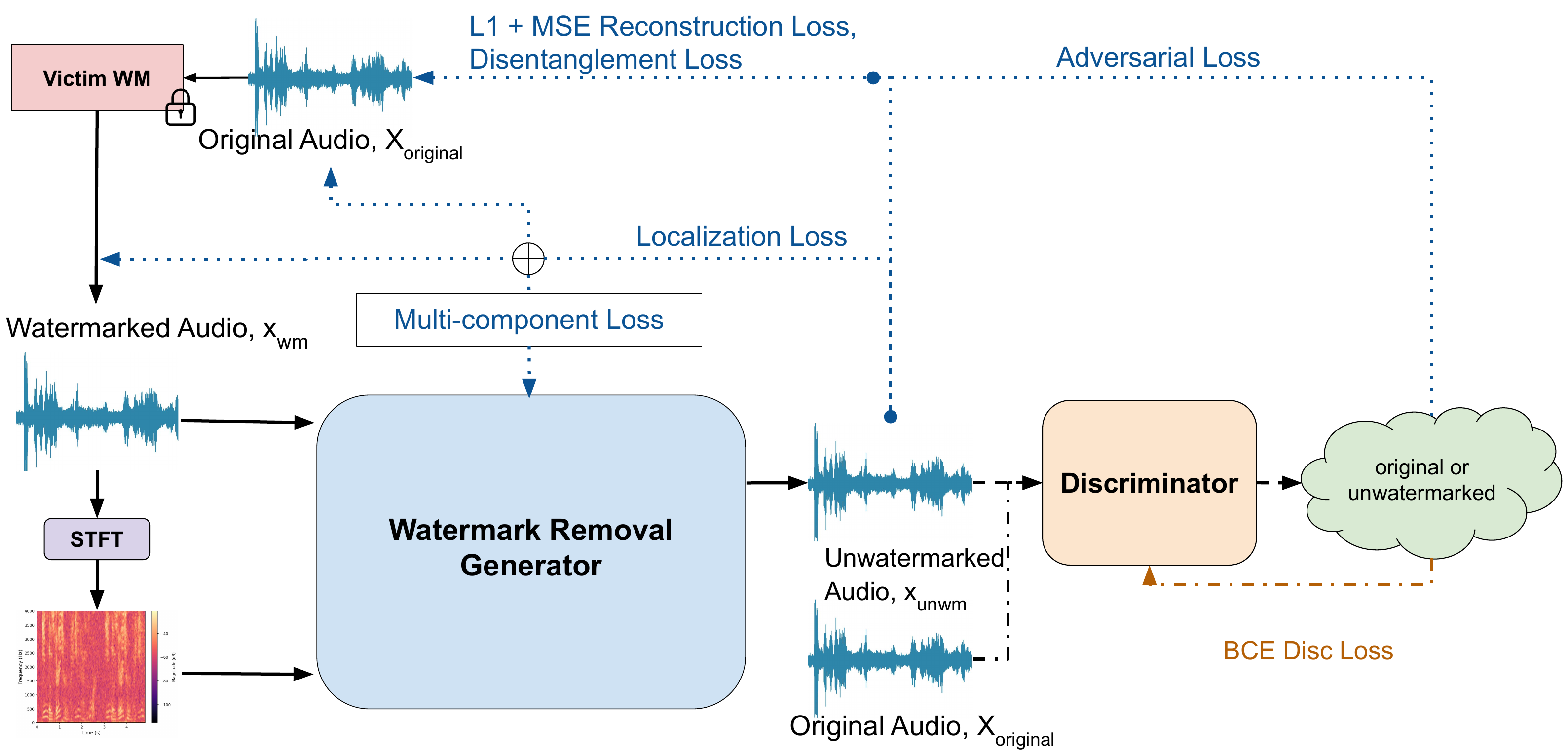}
    \caption{\tool's overview. The approach adopts a dual-path autoencoder architecture for the watermark-removal generator, and a discriminator for GAN-style adversarial training. The watermark-removal generator processes watermarked audio to produce unwatermarked outputs, while the discriminator learns to distinguish these from the corresponding original references. The two models are co-trained iteratively, with the discriminator's feedback guiding the generator towards improved watermark removal and perceptual fidelity.}
    \label{fig:pipeline}
\end{figure*}

\begin{figure*}[ht!]
    \centering
    \includegraphics[width=.9\linewidth]{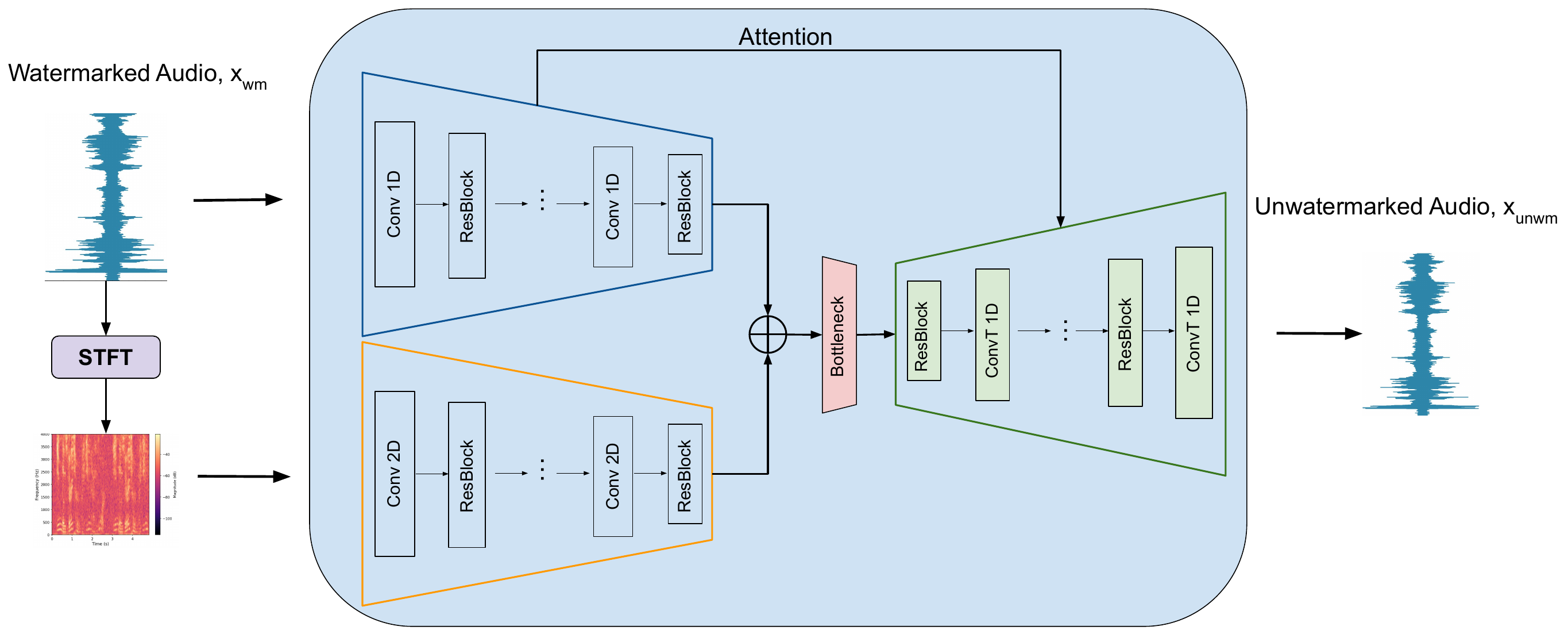}
    \caption{\tool's watermark-removal generator architecture.}
    \label{fig:autoencoder}
\end{figure*}

\section{Design} \label{sec:design}
\subsection{Threat Model and Assumptions}

We study the security of existing audio watermarking schemes from the perspective of an adversary. The target watermarking schemes embed an imperceptible watermark into audio samples and later use a detector to verify watermark presence for provenance, authenticity, or ownership verification. We evaluate whether such watermarks remain detectable. We study an attacker whose goal is to remove watermarks from protected audio such that the target watermark is no longer reliably detected, while the semantic content and perceptual quality are preserved. An attack is considered successful when both criteria are met: (1) a low watermark detection rate on watermark-removed (\emph{unwatermarked}) audio samples; (2) high audio quality of unwatermarked audio samples. 
We consider a black-box attacker. The attacker does not know the watermarking algorithm, detector architecture, detector parameters, training data, secret keys, embedded message, or detection threshold. The attacker also does not require access to the watermark detector during removal. We assume the attacker has access to a dataset of paired audio samples, consisting of original audio samples and their corresponding watermarked versions, for watermark removal model training. The training dataset does not necessarily need to match the distribution or audio type of the watermarked audio produced by the target watermarking scheme. This access assumption captures several realistic attack surfaces. These pairs may be obtained from publicly available sources or leaked content on the internet. A public watermarking or generation service may allow users to submit their own audio and receive watermarked outputs, enabling an attacker to collect input-output pairs before access is rate-limited or revoked. Importantly, the attacker needs such access only during model training; once the removal model is trained, future attacks can be performed offline.

\subsection{Design Overview}
\tool employs adversarial training inspired by GANs~\cite{goodfellow2014generativeadversarialnetworks}: a \textit{watermark-removal generator} that learns to remove watermarks while a \textit{discriminator} is trained simultaneously to distinguish \textit{original} audio from the generator's \textit{unwatermarked} audio outputs. 

The watermark-removal generator is a custom dual-path autoencoder that includes a waveform encoder, a spectrogram encoder, and a decoder. By jointly modeling audio in both time and frequency domains, this architecture enables the model to learn the watermark signal structure and localize watermark artifacts in the time-frequency space, where they are typically embedded in psychoacoustically masked spectrogram bins. These representations are further leveraged by a novel \textit{multi-component loss function}, which allows the model to optimize toward actively opposing the watermark signals while focusing on watermark-prone masked spectrogram bins and maintaining fidelity of the original audio.

Following the adversarial training paradigm, the discriminator distinguishes generated unwatermarked audio from original audio. Both components are optimized jointly: the generator employs the multi-component loss to balance watermark modeling, localization, and perceptual quality preservation, while the discriminator is trained using a standard binary cross-entropy classification loss. 
The system overview is shown in Figure~\ref{fig:pipeline}.

\subsection{Watermark-Removal Generator}

As watermark algorithms may seek to hide their signal in both time and frequency domains, \tool{} uses a novel \textit{dual-path autoencoder} architecture, modifying a standard convolutional autoencoder~\cite{dong2017learningdeeprepresentationsusing} to learn watermark patterns for removal. 
AudioSeal~\cite{audioseal}, for instance, embeds watermarks across multiple frequency bands while maintaining temporal coherence. Single-domain processing (time-only or frequency-only) would fail to capture the full scope of the watermark embedding space. \tool{}'s dual-path autoencoder captures audio watermark information in both the time and frequency domains. As shown in Figure~\ref{fig:autoencoder}, the waveform encoder (top) extracts temporal features from the watermarked audio, while the spectrogram encoder (bottom) captures complementary spectral information from its STFT representation. By jointly modeling these two views, the neural network learns more expressive watermark representations that support both accurate modeling of the watermark structure and localization of watermark signals, which are typically concentrated in psychoacoustically masked time-frequency bins. Their outputs are concatenated in a shared bottleneck layer that learns the joint embeddings, which are then decoded by an attention-enabled decoder that produces unwatermarked audio. 

\newpar{Waveform Encoder}
The \textit{Waveform Encoder}  (Figure~\ref{fig:autoencoder}, top blue trapezoid) processes 1D raw audio signals in the time domain through convolutional layers~\cite{Gradient_based}. Audio samples are standardized to a 16kHz sampling rate. Each convolutional layer incorporates 1D convolutions with decreasing kernel sizes to capture multiscale temporal patterns, along with batch normalization and residual connections to preserve fine-grained temporal information while maintaining training stability. This multiscale feature extraction pipeline captures watermark artifacts at different temporal scales (short-term frames for transient watermark patterns and long-term contexts for more persistent watermark residuals), utilizing the psychoacoustic principle that watermark residuals are embedded across various temporal resolutions to exploit auditory masking~\cite{swansonRobustAudioWatermarking1998, spreadspectral}. This allows \tool to generalize across watermarking schemes that leverage different types of perceptual masking.

\newpar{Spectrogram Encoder}
The \textit{Spectrogram Encoder} (Figure~\ref{fig:autoencoder}, bottom orange trapezoid) applies a Short-Time Fourier Transform (STFT)~\cite{STFT} to the watermarked audio and processes the resulting representation using 2D convolutional layers (as the audio signal becomes 2D data with both time and frequency dimensions). The STFT applied uses a 2048-point FFT with a 512-sample hop length, generating spectrograms spanning 0-8kHz frequency range, which is a setting that balances resolution with efficiency. The spectrogram encoder utilizes the time-frequency domain of audio signals and captures watermark features that might not be detected through temporal analysis alone. Indeed, many audio denoising approaches also operate in the time-frequency domain~\cite{sainburg2020finding, tim_sainburg_2019_3243139}, which inspires our design choice. The architecture mirrors that of the waveform encoder but uses 2D operations for time-frequency representations.

\newpar{Attention-Based Decoder}
In our decoder architecture (Figure~\ref{fig:autoencoder}, right green trapezoid), we enhance the standard convolutional autoencoder design by incorporating attention mechanisms that establish selective connections between corresponding encoder and decoder layers, preserving critical long-range dependencies. Inspired by LightShed's attention-based design~\cite{foersterlightshed}, this architecture enables \tool to focus computational resources on spectrogram bins containing watermark artifacts while preserving the integrity of authentic audio content. As a result, the model achieves more targeted watermark removal during the decoding process.

\subsection{Discriminator}\label{subsec:disc}
Our discriminator is adversarial to the generator. The discriminator distinguishes between original and unwatermarked audio (See Figure~\ref{fig:disc} in Appendix~\ref{app:disc}). It takes a single audio input and outputs a binary classification result.
With the presence of the discriminator, the watermark-removal generator must produce unwatermarked outputs that confuse the discriminator into believing they are original, \ie never watermarked. Therefore, the discriminator provides adversarial supervision forcing the watermark-removal generator to produce outputs indistinguishable from original audio, preventing the application of aggressive noise or distortion that might remove watermarks but significantly degrade audio quality.

\subsection{Loss Function Design}
Our training objective for the watermark-removal generator employs a novel \textit{multi-component loss} function designed to balance watermark removal with perceptual quality. Specifically, our training objective for the generator combines four losses in total, including 
disentanglement loss $\mathcal{L}_{\text{disentangle}}$,
localization loss $\mathcal{L}_{\text{local}}$, reconstruction loss $\mathcal{L}_{\text{recon}}$, 
and adversarial loss $\mathcal{L}_{\text{adv}}$:
\begin{equation*}
\mathcal{L}_{\text{total}} = \alpha_d \mathcal{L}_{\text{disentangle}} + \alpha_l \mathcal{L}_{\text{local}} + \alpha_r \mathcal{L}_{\text{recon}} + \alpha_a \mathcal{L}_{\text{adv}}
\end{equation*}
where $\alpha_d$, $\alpha_l$, $\alpha_r$, and $\alpha_a$ are the tunable hyperparameters that control the relative importance of watermark disentanglement strength, watermark localization awareness, reconstruction quality, and adversarial robustness, respectively.

\newpar{Generator Disentanglement Loss}
The disentanglement loss addresses the first objective (precise removal) of watermark removal: allowing the model to learn the watermark signal structure. It is designed to guide the watermark-removal generator towards explicitly modeling the watermark signal so that it can be effectively removed:

\begin{equation*}
\mathcal{L}_{\text{disentangle}} = \frac{1}{2}\left(1 + \text{cosine\_sim}(\Delta_{\text{proc}}, \Delta_{\text{wm}})\right)
\end{equation*}

where $\Delta_{\text{proc}} = x_{\text{unwm}} - x_{\text{original}}$ and $\Delta_{\text{wm}} = x_{\text{wm}} - x_{\text{original}}$. By minimizing this objective, the model is encouraged to learn the structure of watermark-induced perturbations in a way that enforces guided removal, rather than arbitrary denoising. Specifically, the disentanglement loss provides a direct learning signal that aligns the generator’s modification direction with the inverse of the watermark signal, thereby encouraging the model to learn how the watermark is formed so it can consistently generate opposing transformations that effectively remove it.

\newpar{Generator Localization Loss}
The localization loss addresses the second objective (localization) of watermark removal: identifying where watermark signals are concentrated in the time-frequency domain. It exploits the fundamental principle that audio watermarks are typically embedded in time-frequency bins where the original signal provides natural psychoacoustic masking, and explicitly guides the model to focus on these areas:

\begin{equation*}
\mathcal{L}_{\text{local}} =  \sum_{m=1}^{M} w_m \cdot r_m
\end{equation*}
where the attention weights are computed using a softmax normalization over watermark energy distributions:

\begin{equation*}
w_m = \frac{\exp(e_m)}{\sum_{k=1}^{M} \exp(e_k)}
\end{equation*}
The watermark energy in the mel band m is calculated as the temporal average of power spectral differences:

\begin{equation*}
\begin{aligned}
e_m =
\frac{1}{T}\sum_{t=1}^{T}
\Bigl|
&\mathrm{Mel}_m(\|\mathrm{STFT}(x_{\mathrm{wm}})\|^2)_t \\
&-\mathrm{Mel}_m(\|\mathrm{STFT}(x_{\mathrm{original}})\|^2)_t
\Bigr|
\end{aligned}
\end{equation*}

The processed residual energy represents the remaining artifacts after our removal process:

\begin{equation*}
\begin{aligned}
r_m =
\frac{1}{T}\sum_{t=1}^{T}
\Bigl|
&\mathrm{Mel}_m(\|\mathrm{STFT}(x_{\mathrm{unwm}})\|^2)_t \\
&-\mathrm{Mel}_m(\|\mathrm{STFT}(x_{\mathrm{original}})\|^2)_t
\Bigr|
\end{aligned}
\end{equation*}

Here, $Mel(\cdot)$ represents the mel-scale frequency bands~\cite{pytorchMelScalex2014} covering the perceptually critical range of 200Hz-8kHz, $m$ is an index of the mel frequency band (out of $M$ total bands, each corresponding to a specific perceptual frequency range), and $t$ represents discrete time frames in the STFT analysis window.
This formulation leverages the perceptual structure of the human auditory system, where mel bands reflect non-uniform frequency sensitivity, and aligns optimization with time-frequency bins that are more relevant to watermark embedding.

By operating in this space, the localization loss enables the model to learn where watermark energy is concentrated, rather than treating all frequencies uniformly. This is crucial for achieving targeted watermark removal, as modern watermarking schemes intentionally exploit psychoacoustic masking to hide signals in perceptually insensitive time-frequency bins. Without such localization, optimization would be spread across irrelevant frequency bands, diluting gradients from watermark-intense time-frequency bins and reducing removal effectiveness. In contrast, our localization loss adaptively emphasizes these high-energy masked bins, enabling the model to precisely focus its removal capacity where watermarks are most likely embedded.

\newpar{Generator Reconstruction Loss}
The reconstruction loss addresses the third objective (quality) of watermark removal: preserving the perceptual quality of the underlying audio. It enforces temporal-domain fidelity by directly constraining the waveform-level deviation between the generator’s output and the original audio:
\begin{equation*}
   \mathcal{L}_{\mathrm{recon}}=\|x_{\mathrm{unwm}} - x_{\mathrm{original}}\|_1+\gamma \|x_{\mathrm{unwm}} - x_{\mathrm{original}}\|_2^2 
\end{equation*}
where $x_{\text{wm}}$ denotes the watermarked audio, $x_{\text{original}}$ is the original reference audio, and $x_{\text{unwm}}$ is the generator’s output after watermark removal.
The hybrid formulation combines the robustness of the L1 term to outliers with the smoothness-inducing properties of the L2 term, whose contribution is controlled by $\gamma$.

Unlike localization and disentanglement losses, which focus on where watermark signals reside and how they can be removed, the reconstruction loss operates at the signal level to ensure that the fundamental temporal structure of the audio is preserved. This provides a global constraint that prevents over-aggressive modifications, ensuring that watermark removal does not compromise perceptual quality or introduce audible artifacts.

\newpar{Generator Adversarial Loss} If watermarks are partially removed but the resulting audio exhibits statistical artifacts or unnatural characteristics detectable by machine learning systems, classifiers may still flag these samples as ``unwatermarked'' or ``suspicious'' instead of ``original'' (never watermarked), even without detecting the embedded watermark.
This motivates the inclusion of a discriminator network that enforces distributional realism, thereby enabling a GAN-style adversarial training paradigm.

The adversarial loss encourages the generator to produce audio that is statistically similar to natural, unprocessed audio in the learned feature space of the discriminator. This ensures that even when perfect reconstruction is not achieved by the watermark-removal generator, the unwatermarked audio exhibits the same statistical properties and naturalness as authentic original audio, making it sound more natural to human listeners so that the human perceptual system is less likely to identify processing evidence:

\begin{equation*} \mathcal{L}_{\text{adv}} = \text{BCE}(D(x_{\text{unwm}}), 1)
\end{equation*}

where $D(\cdot)$ represents the discriminator output, BCE denotes binary cross-entropy loss that trains the watermark-removal generator to produce samples that fool the discriminator into classifying them as original audio. the discriminator \(D\) is trained to distinguish original audio from unwatermarked audio, which we discuss in detail in Section~\ref{subsec:disc}. This setup creates a competitive training objective where the generator should produce outputs that are indistinguishable from natural, unprocessed audio.

The adversarial loss complements the other training objectives by ensuring that the unwatermarked audio sounds natural to human ears, even when perfect watermark residual removal is not achieved. 
The adversarial loss optimization objective operates through a learned discriminative signal that is given by the discriminator to the generator, making the optimization complementary to the constraints imposed by other loss components.

\newpar{Discriminator Training Loss}
\label{sec:disc_training}
The discriminator is trained together with the generator using a standard BCE loss that enforces correct classification between two categories, namely original and unwatermarked audio (produced by the generator):

\begin{equation*} \mathcal{L}_{\text{disc}} = \frac{1}{2}\left[\mathcal{L}_{\text{original}} + \mathcal{L}_{\text{unwatermarked}}\right], \end{equation*}
where each component targets a specific audio category:
\begin{align*}
\mathcal{L}_{\text{original}} &= \text{BCE}(D(x_{\text{original}}), 1) \\
\mathcal{L}_{\text{unwatermarked}} &= \text{BCE}(D(G(x_{\text{wm}})), 0)
\end{align*}

Here, $D(\cdot)$ represents the discriminator's classification results, $G(\cdot)$ denotes the watermark-removal generator. The original audio component trains the discriminator to recognize natural, unprocessed original audio (label 1). The unwatermarked audio component ensures the discriminator can detect the watermark-removal generator's outputs (label 0). This formulation creates a comprehensive training signal that enables the discriminator to distinguish original audio from unwatermarked, providing robust adversarial pressure for the watermark-removal generator's improvement.

\subsection{Unified Training and Testing Loop}\label{subsec:unified}
We have two trainable components: (1) watermark-removal generator training with the custom multi-component loss---to establish watermark removal capabilities, and (2) adversarial training with corresponding discriminator training loss and the adversarial loss feeding back to the watermark-removal generator---to achieve robust watermark removal, with respect to the adversarial discriminator. 
After all the models are trained, we test the attack effectiveness and audio quality on unseen test data. The training and testing algorithms are shown in Appendix~\ref{app:algo}.

\section{Evaluation} \label{sec:eval}

In this section, we discuss the experimental evaluation of \tool, focusing on its effectiveness across multiple watermarks and audio domains. We first compare its performance against baseline competitors on different watermark models and data types. We then examine the extent to which \tool transfers to an unseen audio domain without retraining, highlighting its generalizability. Next, we analyze how audio sample length influences the performance of \tool relative to existing methods. Finally, we investigate the contribution of each architectural and loss component to understand which elements have the greatest impact on its watermark removal capability. 

\subsection{Experimental Setup}

\label{sec:eval-setup}
All experiments were performed on a system with two Intel Xeon 6548Y processors, 512GB of RAM, and four Nvidia H100 GPUs having 96GB high-bandwidth memory.

\newpar{Watermarking Methods}
We evaluate \tool against four state-of-the-art audio watermarking schemes: AudioSeal~\cite{audioseal}, AudioMarkNet~\cite{AudioMarkNet}, WavMark~\cite{chenWavMarkWatermarkingAudio2024}, and SilentCipher~\cite{singhSilentCipherDeepAudio2024}. These methods cover diverse watermarking paradigms and are widely evaluated in prior work, making them suitable representative watermarks for our study. 
AudioSeal and SilentCipher are neural audio watermarking schemes that typically employ generator-detector architectures, but they differ in capacity. AudioSeal is a zero-bit scheme whose detector performs binary watermark classification, classifying a sample as watermarked when the confidence score exceeds 50\%. SilentCipher, in contrast, supports multi-bit watermarking by embedding binary strings that can be decoded by its detector. 
WavMark follows a traditional spread-spectrum design, embedding binary string watermarks directly into the frequency domain. AudioMarkNet differs from these post-hoc detection schemes: it embeds watermarks into original speech so that the watermark is learned and transferred through TTS-based voice cloning models. 
Together, our evaluation covers a broad range of audio watermarking paradigms, spanning neural, spread-spectrum, zero-bit, and multi-bit watermarking schemes.

\newpar{Datasets}
Our evaluation employs widely used audio datasets that represent diverse acoustic characteristics and usage scenarios commonly encountered in real-world watermarking applications. 
We utilize two primary datasets: LibriSpeech \cite{librispeech} for speech content, and FMA-small, a subset of Free Music Archive (FMA) \cite{fma_dataset}, for musical content, ensuring comprehensive coverage of different audio domains.
We choose LibriSpeech due to its diverse speaker characteristics, standardized preprocessing, and consistent 16kHz sampling rate. 
The FMA-small dataset, which we refer to as the FMA dataset for brevity, provides musical content with diverse genres, instruments, and acoustic characteristics. It contains 8{,}000 tracks across 8 genres, each with 30 seconds in duration. Music poses unique challenges for watermark removal due to its complex spectral structure with multiple simultaneous instruments that create rich masking opportunities for watermark embeddings.

In our evaluation, AudioMarkNet watermark achieves reliable performance only on its original VCTK speech dataset reported in the paper~\cite{AudioMarkNet}. Since this watermark is designed for speech data, it does not apply to the FMA music dataset. However, even when evaluated on the LibriSpeech speech dataset, it yields a zero watermark detection rate. As a result, we report the evaluation results for AudioMarkNet on the VCTK dataset.

Since several audio watermarking schemes only support or are primarily released for 16kHz audio~\cite{chenWavMarkWatermarkingAudio2024,AudioMarkNet}, we standardize all watermarked and unwatermarked samples to a 16kHz sampling rate.
For our experiments, we use \dtrain to denote the dataset we train the attack on and \deval to denote the dataset we evaluate the attack on. \dtrain and \deval can be different to study \tool transferability (Section~\ref{subsec:transfer}).

\newpar{Baseline Attacks} We compare \tool against three state-of-the-art attacks to demonstrate its effectiveness in attack success, resulting (unwatermarked) audio quality, and processing time.
We select our primary baseline competitor, the best-performing adversarial attack from the recent paper assessing watermarking robustness~\cite {liuAudioMarkBenchBenchmarkingRobustness}, which we refer to as \textit{\audiomarkbench}.
Based on the original image adversarial attack~\cite{andriushchenkoSquareAttackQueryefficient2020}, \audiomarkbench uses optimization to find minimal perturbations that confuse the watermark detector. The attack employs a square-shaped perturbation pattern and iteratively refines it with the watermark detector confidence score by making a large number of queries to the detector. 

Codec compression serves as our second competitor. It represents traditional signal processing attacks through lossy audio compression, as evaluated in recent work~\cite{oreillyDeepAudioWatermarks2025}.
Following this work, we selected multiple codec configurations, including traditional MP3 and OGG, and the neural EnCodec system developed by Meta~\cite{défossez2022highfidelityneuralaudio} at various compression levels. Traditional codecs are tested at bitrates ranging from 32---320 kbps. For traditional codecs, we present the results of the codec/bitrate pair that yields the highest ASR in the experimental results. EnCodec is evaluated at multiple bandwidth settings, and the results are reported for the best-performing configuration. All codecs process audio through a compression-decompression cycle.

Our third baseline is the signal distortion attack~\cite{oreillyDeepAudioWatermarks2025}, which applies a range of signal processing operations to the watermarked audio in an attempt to remove the embedded watermark.

\newpar{Evaluation Metrics}
To evaluate \tool{} and the baseline attacks, we utilize metrics that measure the watermark removal effectiveness and efficiency, and the audio quality preservation capability. The metrics we adopt are as follows:
\begin{enumerate} 
    \item \textbf{Attack Success Rate (ASR)} ($\uparrow$). For zero-bit AudioSeal, we consider a watermark removal successful if the watermark detector's confidence score drops below the detection threshold (\ie  0.5) after processing. For multi-bit watermarks, WavMark considers an attack successful if the decoder fails to extract the payload from the unwatermarked audio. SilentCipher considers an attack successful if the decoded payload message does not match a reference message. AudioMarkNet embeds and decodes a unique watermark payload on each 1-second segment of the test audio. We consider the attack successful if its decoder fails to correctly decode more than 50\% of the segments after removal. ASR is calculated as the proportion of successful watermark removals to the number of processed watermarked samples in total.
    \item \textbf{Virtual Speech Quality Objective Listener (ViSQOL)} ($\uparrow$). ViSQOL evaluates the similarity between watermarked and unwatermarked audio signals using a spectro-temporal comparison inspired by human auditory perception. It has been shown to correlate well with subjective listening tests. Typically, ViSQOL scores range from 0 to 5, where higher values indicate better perceptual quality.
    \item \textbf{Short-Time Objective Intelligibility (STOI)} ($\uparrow$). 
    STOI evaluates speech intelligibility by measuring how well the temporal envelopes of an unwatermarked speech signal match those of the watermarked signal. STOI values range from $0$ to $1$, where values closer to $1$ indicate better preserved intelligibility. STOI is complementary to perceptual quality metrics such as ViSQOL.
    \item \textbf{Signal-to-Noise Ratio (SNR)} ($\uparrow$). SNR measures the signal-level difference between the watermarked and unwatermarked audio. SNR values indicate how close the unwatermarked audio is to a purely signal-level measurement; it does not always correlate with human perception.
    \item \textbf{Attack Time} ($\downarrow$). The attack time metric measures the average time in seconds required to complete the watermark removal process for a single audio sample with our watermark removal model or the baseline methods.
\end{enumerate}

\begin{table}[th!]
\caption{Performance of \tool against baseline competitors on the LibriSpeech dataset. \tool is trained on the LibriSpeech dataset (\dtrain) against AudioSeal and evaluated on unseen samples of the LibriSpeech dataset (\deval), including cross-watermark transfer settings.}
\centering
\setlength{\tabcolsep}{1pt}
\resizebox{\columnwidth}{!}{
\begin{tabular}{llccc|c}
\toprule
\textbf{WM} & \textbf{Attack} & \textbf{STOI} & \textbf{ViSQOL} & \textbf{SNR (dB)} & \textbf{ASR (\%)}\\
\midrule
\multirow{4}{*}{\begin{tabular}{c}SilentCipher\end{tabular}}
  & MP3/OGG         & 0.944 & 3.54 & 1.76  & 100 \\
  & EnCodec         & 0.925 & 4.15 & -0.62 & 100 \\
  & \audiomarkbench & 0.842 & 3.02 & 5.96  & 100 \\
  & \textbf{\tool}            & \textbf{0.939} & \textbf{4.10} & \textbf{11.21} & \textbf{100} \\
\midrule
\multirow{4}{*}{\begin{tabular}{c}WavMark\end{tabular}}
  & MP3/OGG         & 0.971 & 4.39 & 9.72  & 42  \\
  & EnCodec         & 0.921 & 4.03 & 5.92  & 100 \\
  & \audiomarkbench & 0.942 & 4.01 & 17.20  & 84  \\
  & \textbf{\tool}            & \textbf{0.944} & \textbf{4.02} & \textbf{11.32}  & \textbf{100} \\
\midrule
\multirow{4}{*}{\begin{tabular}{c}AudioSeal\end{tabular}}
  & MP3/OGG         & 0.944 & 3.42 & 1.76  & 16  \\
  & EnCodec         & 0.969 & 4.22 & 0.14  & 16  \\
  & \audiomarkbench & 0.941 & 3.94 & 14.46  & 52  \\
  & \textbf{\tool}            & \textbf{0.943} & \textbf{4.03} & \textbf{11.39}  & \textbf{100} \\

\bottomrule
\end{tabular}
}
\label{table-perf-audioseal-librispeech}
\end{table}

\begin{table}[t!]
\caption{Cross-watermark and cross-dataset transfer results of \tool against baseline competitors. \tool is trained on the LibriSpeech dataset against AudioSeal.
The baseline results are collected without watermark transfer or dataset transfer.
}
\centering
\setlength{\tabcolsep}{1pt}

\resizebox{\columnwidth}{!}{
\begin{tabular}{l l c c c | c}
\toprule
\textbf{WM} & \textbf{Attack} & \textbf{STOI} & \textbf{ViSQOL} & \textbf{SNR (dB)} & \textbf{ASR (\%)} \\
\midrule
\multirow{4}{*}{\begin{tabular}{c}SilentCipher \\ (FMA)\end{tabular}}
  & MP3/OGG         & 0.932 &   3.96   &    10.78  & 88  \\
  & EnCodec         & 0.929 &  4.05    &    10.31  & 100 \\
  & \audiomarkbench & 0.886 &   4.02   &   10.89   & 100 \\
  & \textbf{\tool}  & \textbf{0.958} & \textbf{4.39} & \textbf{10.83} & \textbf{100} \\
\midrule
\multirow{4}{*}{\begin{tabular}{c}WavMark \\ (FMA)\end{tabular}}
  & MP3/OGG         & 0.960 &   4.38   &   10.15   & 48  \\
  & EnCodec         & 0.820 &  3.99    &    7.10  & 100 \\
  & \audiomarkbench & 0.890 &   4.24  &   11.88   & 44  \\
  & \textbf{\tool}  & \textbf{0.949} & \textbf{4.31} & \textbf{10.92} & \textbf{100} \\
\midrule
\multirow{4}{*}{\begin{tabular}{c}AudioSeal \\ (FMA)\end{tabular}}
  & MP3/OGG         & 0.926 &   3.93   &   11.98   & 0   \\
  & EnCodec         & 0.928 &   4.26   &   10.40   & 0   \\
  & \audiomarkbench & 0.767 &  3.69    &  6.05    & 2   \\
  & \textbf{\tool}  & \textbf{0.951} & \textbf{4.33} & \textbf{11.21} & \textbf{100} \\
\midrule

\multirow{4}{*}{\begin{tabular}{c}AudioMarkNet \\ (VCTK)\end{tabular}}
  & MP3/OGG         & 0.947 & 3.93 & 18.09  & 1   \\
  & EnCodec         & 0.914 & 3.98 & 5.81  & 34  \\
  & \audiomarkbench & 0.897 & 3.40 & 17.71  & 38  \\
  & \textbf{\tool}  & \textbf{0.930} & \textbf{3.93} & \textbf{11.09}  & \textbf{92}  \\

\bottomrule
\end{tabular}
}
\label{table-perf-fma-transfer}
\end{table}

\subsection{Performance Evaluation}
\label{sec:eval-transfer}
We train our universal watermark removal model on the LibriSpeech dataset against the AudioSeal watermarking scheme. 
Table~\ref{table-perf-audioseal-librispeech} shows its performance against the baselines on unseen samples in LibriSpeech, watermarked with SilentCipher, WavMark, and AudioSeal, where \tool's results reported for SilentCipher and WavMark are \emph{cross-watermark} transfer results. 
Then, we evaluate the \emph{cross-dataset} performance of \tool's universal watermark removal model on the FMA dataset against AudioSeal and simultaneously \emph{cross-watermark/cross-dataset} performance on the FMA dataset against SilentCipher and WavMark.  
Since AudioMarkNet does not support LibriSpeech and FMA (Section~\ref{sec:eval-setup}), we evaluate \tool's \emph{cross-watermark/cross-dataset} performance on the VCTK dataset used in the original AudioMarkNet paper~\cite{AudioMarkNet}. The results of this evaluation are presented in Table~\ref{table-perf-fma-transfer}. It is important to note that, because the competing removal methods are not \lbased, their results do not reflect cross-watermark or cross-dataset transfer; rather, they correspond to independent evaluations on each target watermark and dataset combination.

\newpar{Attack Success Rate} Tables~\ref{table-perf-audioseal-librispeech} and \ref{table-perf-fma-transfer} show that some watermarking schemes are highly vulnerable to removal. SilentCipher, for example, can be almost consistently removed by existing attacks across LibriSpeech and FMA, with all attacks achieving 100\% ASR except for MP3/OGG. However, more robust watermarks exist. For instance, against WavMark, MP3/OGG and \audiomarkbench achieve only 42\% and 84\% ASR on LibriSpeech, respectively, and both drop below 50\% ASR on FMA. Furthermore, when evaluated against AudioSeal, MP3/OGG and EnCodec fail on both datasets, while \audiomarkbench barely reaches 50\% ASR on LibriSpeech and only 2\% ASR on FMA. In contrast, \tool achieves 100\% ASR across SilentCipher, WavMark, and AudioSeal on both LibriSpeech and FMA datasets, even though it is only trained on LibriSpeech samples against AudioSeal. 
A similar trend is observed when testing on the VCTK dataset against AudioMarkNet, the highest ASR achieved by the baseline attacks is 38\% (\audiomarkbench), substantially below \tool{}'s 92\% ASR. The results demonstrate that \tool is not only effective against the watermark and the dataset it is trained on, but also transfers across watermarking methods on unseen datasets with different data distributions.

Together with the attacks reported in Tables~\ref{table-perf-audioseal-librispeech} and \ref{table-perf-fma-transfer}, we evaluated eight diverse signal distortion attacks described in~\cite{oreillyDeepAudioWatermarks2025}. As shown in Table~\ref{tab:signal_editing} in Appendix~\ref{app:signal}, this baseline consistently shows a low ASR across the LibriSpeech and FMA datasets, with most distortions achieving zero ASR on FMA samples and near-zero ASR on LibriSpeech, with the best-case lowpass filter reaching merely 10\%. This demonstrates the limitation of such approaches, as their blind, non-adaptive processing fails to reliably identify and remove the embedded watermark.

\begin{figure*}[ht!]
    \centering

    \subfloat[Watermarked audio spectrogram with red boxes indicating watermark regions.]
    {
        \includegraphics[width=0.31\textwidth]{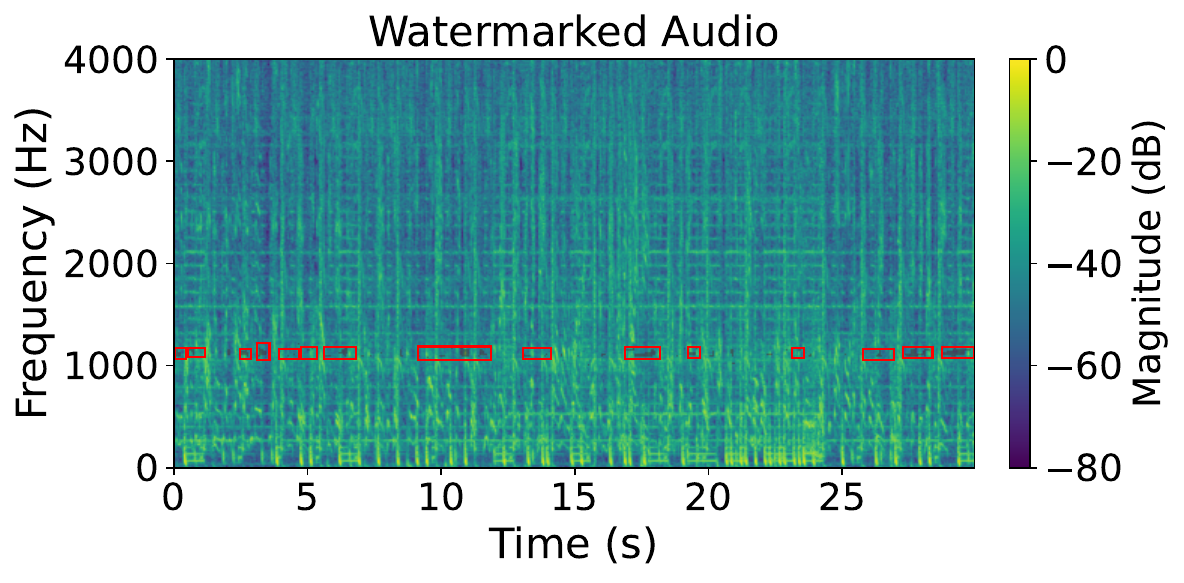}
        \label{fig:watermark_spec}
    }\hfill
    \subfloat[\tool{}-processed audio showing watermark removal regions in blue.]
    {
        \includegraphics[width=0.31\textwidth]{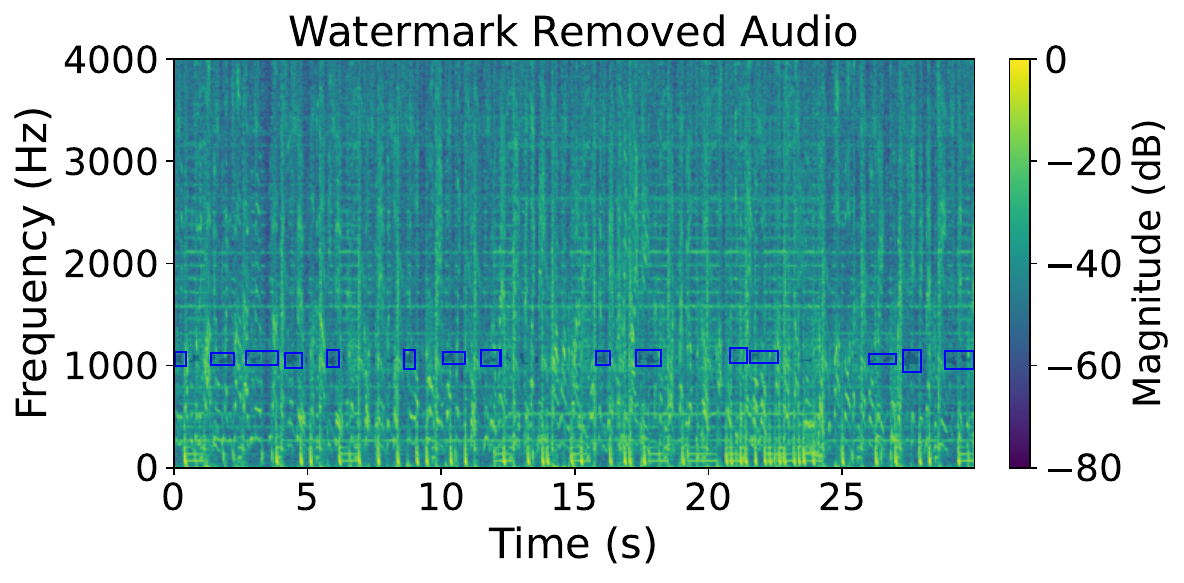}
        \label{fig:harmonicaudio_remove_diff}
    }\hfill
    \subfloat[\audiomarkbench-processed audio with corresponding removal regions.]
    {
        \includegraphics[width=0.31\textwidth]{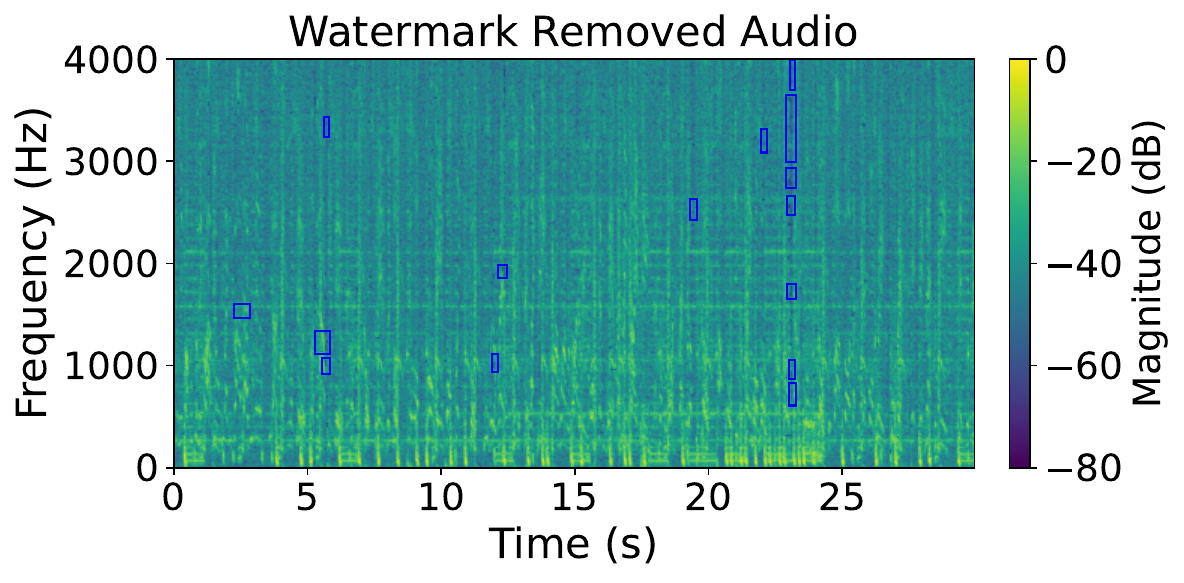}
        \label{fig:audiomarkbench_remove_spec}
    }

    \caption{Comparison of spectrograms for watermarked audio, \tool{} removal, and \audiomarkbench removal on FMA AudioSeal sample. \tool is evaluated by transferring the model trained on AudioSeal LibriSpeech samples.}
    \label{fig:three_spectrograms}
\end{figure*}

\begin{figure*}[ht!]
    \centering

    \subfloat[Watermarked audio spectrogram with red boxes indicating watermark regions.]
    {
        \includegraphics[width=0.31\textwidth]{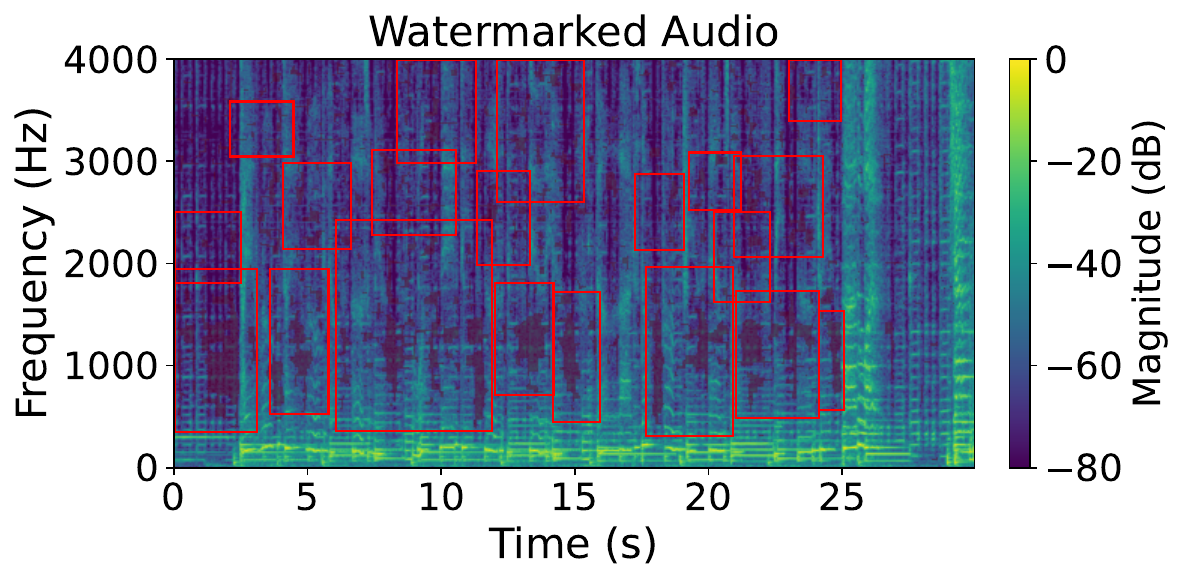}
        \label{fig:WavMark_watermark_spec}
    }\hfill
    \subfloat[\tool{}-processed audio showing watermark removal regions in blue.]
    {
        \includegraphics[width=0.31\textwidth]{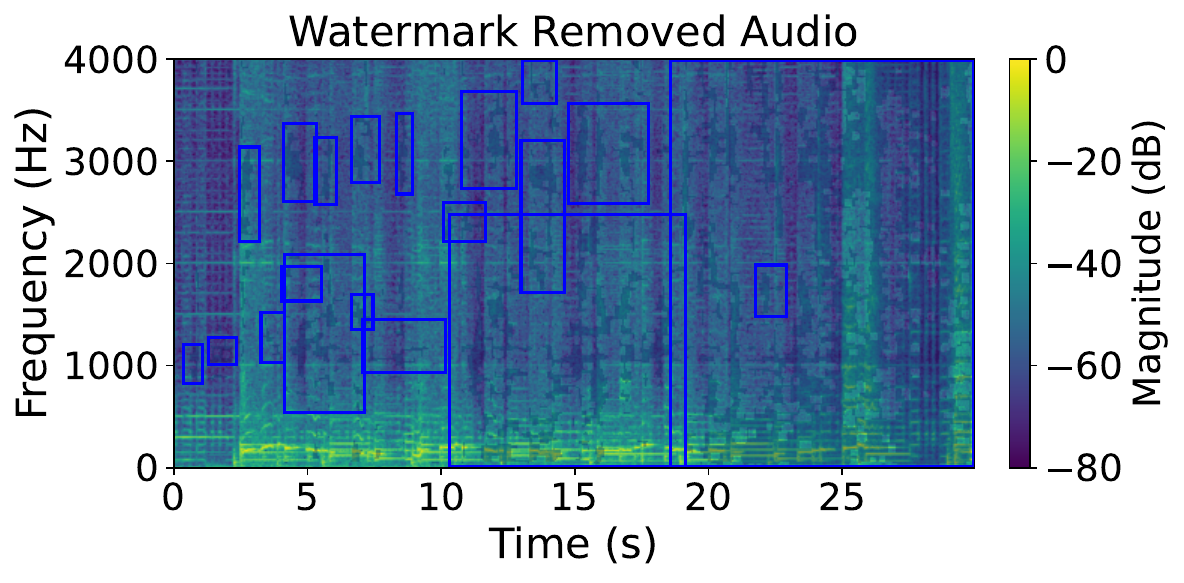}
        \label{fig:WavMark_harmonicaudio_remove_diff}
    }\hfill
    \subfloat[\audiomarkbench-processed audio with corresponding removal regions.]
    {
        \includegraphics[width=0.31\textwidth]{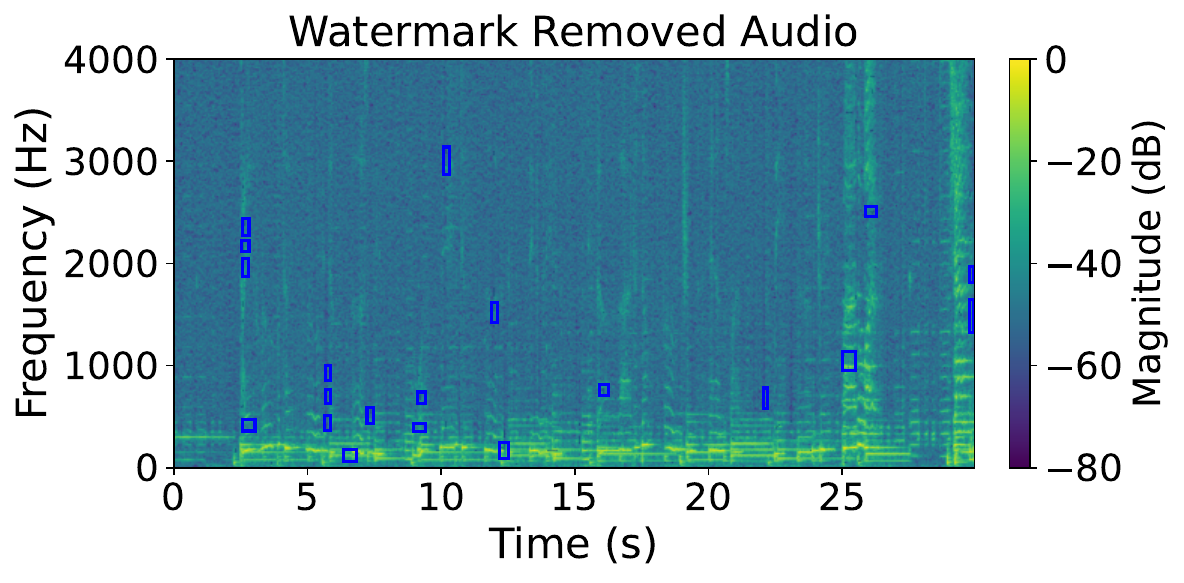}
        \label{fig:WavMark_audiomarkbench_remove_spec}
    }

    \caption{Comparison of spectrograms for watermarked audio, \tool{} removal, and \audiomarkbench removal on FMA WavMark sample. \tool is evaluated by transferring the model trained on AudioSeal LibriSpeech samples.}
    \label{fig:WavMark_three_spectrograms_WavMark}
\end{figure*}

\newpar{Perceptual Quality} A successful removal attack should remove the watermark while preserving the quality of the resulting audio. In Tables~\ref{table-perf-audioseal-librispeech} and \ref{table-perf-fma-transfer}, we report three quality metrics: STOI, ViSQOL, and SNR.

The high STOI scores indicate that speech or vocals remain highly intelligible after the attack, allowing listeners or downstream speech systems to recover the words or lyrics. This metric is more informative for speech datasets than for music datasets.
The high ViSQOL scores further indicate that the resulting audio retains strong perceptual quality, with limited audible degradation relative to the reference audio. In all evaluation settings, \tool achieves high perceptual quality in terms of STOI and ViSQOL while maintaining near-perfect ASR. For instance, on LibriSpeech, \tool obtains STOI scores above 0.939 and ViSQOL scores above 4.0 for all three watermarking methods with 100\% ASR. Other baseline attacks achieve similar perceptual quality with generally lower and less consistent performance in ASR. In those cases where the baseline attack has slightly higher STOI and ViSQOL values, they achieve lower ASR, \eg MP3/OGG's 42\% ASR against WavMark and EnCodec's 16\% ASR against AudioSeal. It is worth noting that we consistently achieve comparable or higher STOI and ViSQOL values than the most competitive attack, \audiomarkbench, with consistently higher ASR.
On FMA and VCTK, \tool similarly achieves the best or near-best STOI and ViSQOL scores across all the evaluated watermarks, demonstrating that its cross-dataset and cross-watermark removal success does not come from damaging the audio signal.

We provide SNR results, but note that SNR should be interpreted with caution as a quality metric for audio intended for human consumption. SNR measures signal-level similarity between watermarked and unwatermarked audio, but it does not necessarily align with perceived quality. In fact, ITU-R BS.1387 explicitly notes that conventional objective measures (\eg SNR ratio and distortion) are no longer adequate for assessing perceived audio quality in modern audio processing systems~\cite{itu2023bs1387}. This limitation arises because SNR is an energy-based metric that does not account for psychoacoustic effects such as frequency-dependent sensitivity, temporal masking, and spectral masking~\cite{painter2000perceptual,brandenburg2013perceptual}. Our measurements show that compared to other baselines,  \tool{} is often able to preserve both perceptual quality and SNR, but when it is not able to, \tool{}'s loss function prioritizes the perceptual metrics.

\subsection{Transferability Analysis}\label{subsec:transfer}

In this section, we analyze \tool's and \audiomarkbench's behavior during the evaluation to identify the reasons why \tool demonstrates strong transferability across various audio data types and watermarks. 
In particular, \tool's removal model trained only on AudioSeal-watermarked LibriSpeech data transfers to VCTK and FMA data watermarked by different watermarks such as AudioMarkNet and WavMark. In contrast, although \audiomarkbench does not need to train a model, its performance is largely dependent on the target audio data type.

We attribute this strong generalization capability to the ability of \tool to jointly learn how to model watermark signals, and where they are located, enabled by several architectural and objective-oriented design choices. First of all, our multiscale resolution encoder captures both fine-grained temporal patterns (crucial for speech) and coarse-grained structures (crucial for music). This supports effective modeling and thus the removal of watermark signals without requiring domain-specific retraining. Second, the attention-augmented decoder dynamically focuses on spectrally anomalous time-frequency bins, regardless of the underlying audio content type. Rather than relying on fixed perturbation patterns, it adaptively identifies regions that deviate from the natural audio distribution. Unlike \audiomarkbench's fixed adversarial perturbations, our attention mechanism learns to identify watermark-induced deviations in the time-frequency space, enabling precise localization of watermark signals across domains. Third, our localization loss, with frequency-adaptive weighting, exploits the fact that many watermarking schemes embed signatures in perceptually less sensitive time-frequency bins or bins where the acoustic masking effect~ \cite{swansonRobustAudioWatermarking1998, spreadspectral} is strong. By emphasizing these bins during optimization while maintaining reconstruction constraints, \tool achieves targeted and localized removal while preserving perceptual fidelity.

To confirm our intuition, we perform a spectrogram analysis.
First, we select an original FMA sample and its corresponding AudioSeal-watermarked sample and locate where the watermarks live. Then, to identify whether \tool and \audiomarkbench target these watermarked bins, we compare the spectrograms of the watermarked and the corresponding unwatermarked audio.
Figures \ref{fig:watermark_spec}, \ref{fig:harmonicaudio_remove_diff}, and \ref{fig:audiomarkbench_remove_spec} show time-frequency spectrograms generated using STFT analysis with normalized magnitude values displayed in dB scale (0 to -80 dB range), with highlighted regions indicating watermark locations and removal patterns overlaid on the spectral representations.
These spectrograms demonstrate that \tool successfully localizes watermark-dense time-frequency areas and applies targeted modifications to remove them. In contrast, \audiomarkbench removes signals spread across the whole time-frequency space and fails to focus on the watermark-relevant spectrogram bins. 
This explains why \audiomarkbench performs well on speech samples but fails on music samples: the search space for music is significantly larger than that of speech, as music often contains multiple instruments and vocal tracks spanning diverse frequency bands. This results in substantially more complex and heterogeneous masking conditions and, consequently, more complicated watermark patterns, making it difficult for fixed perturbation strategies to generalize effectively. Therefore, \audiomarkbench's adversarial search becomes exponentially more challenging in the complex spectral landscape of music with overlapping instrumental harmonics and concurrent melodic lines. Without explicit mechanisms for watermark modeling and localization, its adversarial search becomes inefficient in such settings.
For completeness, we further provide the spectrograms of the watermark signal before and after removal by \tool and \audiomarkbench in Figure~\ref{fig:three_difference_spectra} in Appendix \ref{appendix:watermark_remove_diff}.

We also analyze the spectrograms of WavMark-watermarked audio samples presented in Figures \ref{fig:WavMark_watermark_spec}, \ref{fig:WavMark_harmonicaudio_remove_diff}, and \ref{fig:WavMark_audiomarkbench_remove_spec}. The corresponding spectrograms of watermark signals are in Figure~\ref{fig:three_difference_spectra_wavmark} in Appendix \ref{appendix:watermark_remove_diff}. These spectrograms provide two key observations supporting the transferability discussion of \tool{}, across different datasets and different watermarking schemes. First, although watermarking schemes use distinct embedding heuristics, they are all constrained by and exploit the auditory masking effect \cite{swansonRobustAudioWatermarking1998, spreadspectral}. Rather than concentrating on a single frequency, watermark residuals are embedded within perceptually masked time-frequency components determined by the underlying audio content. As a result, while the exact embedding patterns differ, these masked components exhibit consistent structure and partial overlap across watermarking schemes. This shared property enables \tool{}, trained solely on AudioSeal, to generalize to unseen schemes such as WavMark by learning to both identify and remove watermark perturbations in these time-frequency bins. Second, WavMark's watermark residuals are spread much more broadly across the time-frequency domain, whereas AudioSeal produces more localized residuals. This phenomenon arises because AudioSeal employs a time-frequency loudness loss (TF-Loudness) that concentrates watermark energy in narrow, perceptually optimal locations. In contrast, WavMark utilizes invertible neural networks that operate across the entire spectrogram, spreading watermark information more broadly across frequency and time dimensions. 
Although \audiomarkbench is consistently weaker than \tool against both AudioSeal and WavMark on FMA, its use of random square-noise perturbations increases the chance of overlapping WavMark’s more widely distributed watermark patterns. Consequently, \audiomarkbench achieves stronger ASR against WavMark than AudioSeal, as shown in Table~\ref{table-perf-fma-transfer}.

\subsection{Impact of Audio Sample Length}

\begin{table}
\caption{Impact of FMA sample length on \tool's effectiveness compared to \audiomarkbench. \tool is trained on LibriSpeech and evaluated on FMA.}
\centering
\small
\setlength{\tabcolsep}{1pt}
\begin{tabular}{l l c c c c | c}
\toprule
\textbf{Len (s)} & \textbf{Method} & \textbf{STOI} & \textbf{ViSQOL} & \textbf{SNR (dB)} & \textbf{Time (s)} & \textbf{ASR (\%)} \\
\midrule
\multirow{2}{*}{1}
& \audiomarkbench & 0.764 & 3.56 & 5.55 & 18.07 & 38 \\
& \textbf{\tool}   &\textbf{ 0.952} &\textbf{ 4.27} & \textbf{11.56 }& \textbf{0.019} & \textbf{100} \\
\midrule
\multirow{2}{*}{5}
& \audiomarkbench & 0.768 & 3.54 & 5.74 & 44.39 & 12 \\
& \textbf{\tool}   & \textbf{0.953} & \textbf{4.32} & \textbf{11.39} & \textbf{0.035} & \textbf{100} \\
\midrule
\multirow{2}{*}{10}
& \audiomarkbench & 0.759 & 3.59 & 5.76 & 69.01 & 8 \\
& \textbf{\tool}   & \textbf{0.952 }& \textbf{4.32} & \textbf{11.18} & \textbf{0.037} & \textbf{99} \\
\midrule
\multirow{2}{*}{15}
& \audiomarkbench & 0.758 & 3.59 & 5.84 & 78.48 & 6 \\
& \textbf{\tool}   & \textbf{0.951} & \textbf{4.33} & \textbf{11.14} & \textbf{0.031} & \textbf{99} \\
\midrule
\multirow{2}{*}{30}
& \audiomarkbench & 0.767 & 3.69 & 6.05 & 102.28 & 2 \\
& \textbf{\tool}   & \textbf{0.951} &\textbf{ 4.33} & \textbf{11.21} &\textbf{ 0.029} & \textbf{100} \\
\bottomrule
\end{tabular}
\label{tab:length_comparison}
\end{table}
The results in Table~\ref{tab:length_comparison} demonstrate how audio sample length affects the performance of \tool versus \audiomarkbench when attacking AudioSeal watermarks on FMA samples. 
\tool achieves consistently high ASR (99--100\%) while preserving strong perceptual audio quality, with ViSQOL consistently above 4.2. Moreover, it maintains near-real-time inference across all evaluated audio durations, requiring only around 0.03 seconds per sample. In contrast, \audiomarkbench suffers substantial performance degradation as the audio length increases: its ASR decreases from 38\% on 1-second clips to only 2\% on 30-second samples. It also yields poorer audio quality, with ViSQOL scores around 3.5. Additionally, the computational cost of \audiomarkbench increases with sample length, with attack time rising by 566\%, from 18 seconds for 1-second audio to more than 102 seconds for 30-second samples.

This length-dependent degradation shows a fundamental limitation of \audiomarkbench. Its reliance on random perturbations applied across the entire time-frequency space prevents it from identifying and targeting watermark-bearing bins, especially as audio length and spectral complexity increase. As the search space grows, successful watermark removal becomes increasingly unlikely. 

In contrast, \tool maintains stable performance across varying lengths by explicitly addressing all three objectives of watermark removal. First, for precise removal, it learns to model and remove watermark perturbations, enabling consistent ASR regardless of input duration. Second, its localization objective focuses on watermark-relevant time-frequency bins, avoiding the substantial growth of the search space. Third, its quality objective ensures that these modifications preserve the perceptual structure of the audio, maintaining high quality even for long samples.

This scalability is critical for real-world scenarios, where audio content often spans minutes or hours (e.g., music tracks, podcasts, or streaming audio). Furthermore, due to its high computational overhead, \audiomarkbench cannot support real-time applications such as live voice conversion, where latency must be within tens of milliseconds~\cite{llvc2023,dualvc32024,rtvc2025,yang2024streamvcrealtimelowlatencyvoice}. Prior work in psycholinguistics shows that conversational turn-taking occurs around 200\,ms, while delays above 700\,ms degrade interaction quality~\cite{timingpsychology, castillolopezetal2025survey}. Therefore, \audiomarkbench{} is impractical for realistic deployment. In contrast, \tool{}’s consistent watermark removal performance and near-constant inference time make it a practical threat model for real-world watermark removal.

\begin{table}[th!]
\caption{Ablation study of the effect of the discriminator component on \tool's performance on the LibriSpeech dataset against AudioSeal watermark.}
\centering
\setlength{\tabcolsep}{3pt}
\begin{tabular}{l c c c @{\hspace{6pt}} | @{\hspace{6pt}} c}
\toprule
\textbf{Ablation} & \textbf{STOI} & \textbf{ViSQOL} & \textbf{SNR (dB)} & \textbf{ASR (\%)} \\
\midrule
HA w/o discriminator   & 0.882 & 3.47 & -0.83 & 71 \\
\textbf{\tool} & \textbf{0.943} & \textbf{4.03} & \textbf{11.39} & \textbf{100} \\
\bottomrule
\end{tabular}
\label{tab:ablation}
\end{table}

\begin{table}[th!]
\caption{Ablation study of \tool's multi-objective loss function evaluated on LibriSpeech against AudioSeal.}
\centering
\setlength{\tabcolsep}{1pt}
\small
\begin{tabular}{l c c c @{\hspace{6pt}} | @{\hspace{6pt}} c}
\toprule
\textbf{Ablation} & \textbf{STOI} & \textbf{ViSQOL} & \textbf{SNR (dB)} & \textbf{ASR (\%)} \\
\midrule
$\mathcal{L}_{\text{local}} + \mathcal{L}_{\text{recon}}$& $0.848$ & $3.95$ & $5.097$ & 64 \\
$\mathcal{L}_{\text{disentangle}} + \mathcal{L}_{\text{recon}}$ & $0.970$ & $4.35$ & $10.44$ & 75 \\
$\mathcal{L}_{\text{disentangle}} + \mathcal{L}_{\text{local}}$ & $0.751$ & $3.58$ & -1.53 & 96 \\
\midrule
\textbf{All ($\mathcal{L}_{\text{disentangle}} + \mathcal{L}_{\text{local}} + \mathcal{L}_{\text{recon}}$)} & \textbf{0.943} & \textbf{4.03} & \textbf{11.39} & \textbf{100} \\
\bottomrule
\end{tabular}
\label{tab:ablation_loss}
\end{table}

\subsection{Ablation Study}
\newpar{Effect of Discriminator}
In order to study the effect of the discriminator component in \tool's architecture  (Figure~\ref{fig:pipeline}), we conduct an ablation study.
Specifically, we evaluate our design in the following settings: 
\begin{enumerate}
    \item \textit{HA w/o discriminator:} Watermark-Removal Generator only
    \item \textit{\tool:} Watermark-Removal Generator and Discriminator --- this is our main setting.
\end{enumerate}
For this set of experiments, we set the training dataset \dtrain and the testing dataset \deval to LibriSpeech. The audio is watermarked with AudioSeal.

The results are presented in Table~\ref{tab:ablation}.
As expected, incorporating the discriminator component increases the ASR from 71\% to 100\%, demonstrating that the discriminator-guided generator can remove embedded watermarks more effectively.
At the same time, \tool achieves higher perceptual quality than \tool without the discriminator in terms of STOI, ViSQOL, and SNR.
This improvement highlights the role of the discriminator in enforcing the quality objective, preserving perceptual quality, by encouraging the generator to produce outputs that are indistinguishable from natural audio, while still effectively removing watermark artifacts.

\newpar{Effect of Each Loss Component}
To understand the contribution of each term in \tool's multi-objective loss function, we perform an ablation study by selectively removing individual loss components while keeping all other training settings fixed.
The experiments are conducted on LibriSpeech, where the model is trained against AudioSeal on LibriSpeech. 

Table~\ref{tab:ablation_loss} summarizes the results. They illustrate that excluding $\mathcal{L}_{\text{disentangle}}$ significantly degrades both ASR (64\%) and perceptual quality, demonstrating that explicitly modeling the structure of the watermark signal (the precise removal objective) is critical for effective removal. Without this component, the model lacks a direct learning signal to infer and thereby remove the watermark, leading to weaker and less stable removal behavior.
The exclusion of $\mathcal{L}_{\text{local}}$, which is responsible for identifying where watermark energy is concentrated in the time-frequency space, reduces ASR to 75\% while slightly improving perceptual quality. This indicates that without accurate localization (the localization objective), the model fails to locate all watermark-carrying time-frequency bins, resulting in higher audio fidelity due to less removal effort.
Dropping $\mathcal{L}_{\text{recon}}$ component slightly reduces the ASR from 100\% to 97\% and causes a notable drop in perceptual quality (\ie STOI drops from 0.943 to 0.751), indicating its role in enforcing the quality objective of preserving perceptual fidelity during watermark removal.

Overall, these results confirm that the three loss components directly correspond to the three key objectives of watermark removal: $\mathcal{L}_{\text{disentangle}}$ enables the model to learn watermark structure, $\mathcal{L}_{\text{local}}$ guides the model to further focus on watermark-concentrated time-frequency bins, and $\mathcal{L}_{\text{recon}}$ preserves perceptual quality. Their combination is necessary to achieve both a high attack success rate and high-fidelity unwatermarked audio.

\section{Discussion} \label{sec:discuss}

\subsection{Defenses against \tool}
\tool is the first \lbased attack demonstrating that modern audio watermarks are susceptible to removal in realistic black-box scenarios.
Thus, there is a need for research to design effective watermarks capable of counteracting such removal.

From our empirical analyses of spectrograms before and after watermark removal, we observe distinct time-frequency bins where schemes inject watermarks and where our attack successfully suppresses watermark-related signal perturbations. These findings indicate that the spectrogram representation preserves rich structural and statistical information, offering an expressive space for watermark embedding. Building upon this insight, a promising line of defense is to leverage the time-frequency domain by transforming audio signals into spectrograms and adapting the ideas derived from advanced image watermarking schemes to this domain.
Image watermarking has been studied more extensively than audio watermarking, and has latent-space and semantic watermarking schemes that achieve high robustness against manipulations~\cite{Wang2024_SleeperMark, wen2023treeringwatermarksfingerprintsdiffusion, Ci2024_RingID}.
Such cross-modal adaptation has the potential to integrate the robustness of image watermarking into the audio modality.

\subsection{Limitations}
\tool demonstrates strong transferability across a range of state-of-the-art watermarking schemes due to shared algorithmic principles, 
\ie~watermark perturbations follow perceptual masking rules~\cite{swansonRobustAudioWatermarking1998, spreadspectral}. As observed in the spectrogram analysis in Section~\ref{subsec:transfer}, a model trained on AudioSeal effectively transfers to WavMark because a portion of WavMark's watermark-embedded time-frequency bins overlap with those of AudioSeal, allowing \tool to exploit similar spectral and temporal patterns for watermark removal. However, such generalization is not guaranteed for all possible watermarking schemes and data samples, especially in scenarios where degrading the protected audio quality is acceptable. In such cases, additional finetuning on samples from the new scheme would be required to adapt the model effectively.

\section{Related Work} \label{sec:rw}
\subsection{Audio Watermarking} 
\label{sec:rw:watermark}

Recent advances in generative audio models have made synthetic content nearly indistinguishable from human-created works, raising growing concerns about provenance, copyright, and authenticity. To address these issues, researchers have developed watermarking techniques that embed imperceptible but detectable signals. These watermarking schemes can be categorized into multi-bit, which embed binary messages that the detector can decode, and zero-bit, in which the detector operates as a binary classifier that determines whether a watermark is present in the audio.

Among the zero-bit watermarks, there is AudioSeal~\cite{audioseal}, which represents a major step toward localized watermarking. It jointly trains a generator-detector pair where the generator produces an additive watermark waveform, and the detector estimates probabilities of watermark presence in each sample segment. The model is optimized under a combination of perceptual and detection losses. This setup yields watermarks that are both imperceptible and robust to common signal editions. Building upon AudioSeal, Latent Watermarking of Audio Generative Models~\cite{romanLatentWatermarkingAudio2025} extends watermarking from the waveform level to the latent space of audio language models. Unlike post-hoc watermarking, this method embeds watermarks into the training data of the audio model itself, ensuring that any model trained on such data produces watermarked outputs. Audio Watermark~\cite{guoAudioWAtermArkDynamic} introduces a dynamic, style-transfer-based watermarking framework designed for black-box voice dataset ownership verification. By leveraging out-of-domain features and bi-level adversarial optimization, it produces harmless watermarks that preserve the label while maintaining high verification accuracy.

Among the multi-bit watermarks, WavMark~\cite{chenWavMarkWatermarkingAudio2024} reframes watermarking as a reversible transformation task using invertible neural networks (INNs). It encodes binary messages into spectrograms via STFT and jointly processes audio and watermark features through INN blocks to ensure watermark detection. Timbre Watermarking~\cite{liuDetectingVoiceCloning2023} embeds watermark information directly into the frequency domain of a speaker's voice, leveraging repeated embedding to enhance robustness against common audio preprocessing. GROOT~\cite{liuGROOTGeneratingRobust2024} introduces a generative audio watermarking paradigm that embeds watermarks directly within diffusion-based vocoders by injecting a learned latent watermark vector into the model's input noise space. By jointly training an encoder-decoder around a frozen diffusion model, GROOT enables plug-and-play watermark synthesis. AudioMarkNet~\cite{AudioMarkNet} embeds watermarks into source speech so that the watermark can be learned and transferred through downstream TTS-based voice-cloning models, enabling deepfake speech detection after downstream model finetuning.

\subsection{Audio Watermark Removal}
\label{sec:bg:removalAttack}

Watermarking schemes are susceptible to removal attacks that can erase the embedded patterns. The existing attacks can be categorized into signal-processing-based attacks and \obased attacks based on the methodology.

Signal-processing-based attacks rely on conventional signal processing techniques, such as bandpass filtering or lossy compression codecs, to manipulate watermarked samples with the goal of distorting or suppressing embedded watermark patterns. Prior studies~\cite{oreillyDeepAudioWatermarks2025, ozerComprehensiveRealWorldAssessment2025} have systematically evaluated the effectiveness of these transformations against a range of watermarking schemes. However, such approaches often fail to fully eliminate modern, robust watermarks without introducing noticeable or audible artifacts that degrade overall audio quality.

Optimization-based attacks form a more advanced class of watermark removal methods in which the attack perturbations are explicitly optimized using feedback from the watermark detector's confidence scores. In contrast to signal-processing attacks that apply fixed transformations, these approaches craft a sample-specific perturbation tailored to each watermarked input, aiming for stronger removal and better perceptual quality. Liu et al.~\cite{liuAudioMarkBenchBenchmarkingRobustness} survey a range of such methods, from square attacks~\cite{andriushchenkoSquareAttackQueryefficient2020} to adversarial attacks that rely on full knowledge of the detector's parameters. These techniques assume different levels of prior knowledge of the watermarking scheme, such as access to the detector confidence score or even access to the watermark model's internal parameters. For fairness, we compare against the strongest method presented in the survey---the square attack---which stays within our threat model, although it still operates under stronger assumptions by requiring repeated queries to the watermark detector.

\section{Conclusion} \label{sec:conclusion}
In this work, we introduce \tool, the first \lbased audio watermark removal attack designed to operate under realistic black-box settings. Unlike prior approaches that rely on watermark detector access, extensive querying, or sample-specific optimization, \tool learns a generalizable removal model that effectively removes watermarks while preserving high audio quality.
At the core of our approach is a unified design that addresses three key objectives of modern watermark removal: learning the structure of watermark perturbations to enable their removal (precise removal), identifying where watermark energy is concentrated in the time-frequency domain (localization), and preserving the fidelity of the underlying audio (quality). By jointly integrating a dual-path architecture with a multi-objective loss, \tool achieves strong removal performance without introducing perceptible artifacts.

Empirical results demonstrate that \tool consistently neutralizes state-of-the-art watermarking schemes, achieving near-perfect ASR, near-real-time inference, and high perceptual quality across both speech and music, including on longer audio samples. In contrast, existing attacks either fail to effectively remove modern watermarks, degrade in performance with increasing audio length, or incur substantial computational overhead.
Overall, \tool reveals that modern audio watermarking schemes remain vulnerable when these watermarking removal objectives are jointly addressed. Our findings highlight the need for future watermarking designs that move beyond robustness to simple perturbations and instead account for adaptive, learning-based attacks that explicitly model, localize, and suppress watermark signals.

\bibliographystyle{ACM-Reference-Format}
\bibliography{ref}

\appendix %
\section*{Open Science} %
This research complies with the ACM CCS Open Science policy. Our artifact is available through the following anonymized link: \url{https://anonymous.4open.science/r/harmonicaudio-B18A/}
. It includes the materials needed to evaluate the paper’s contributions, including code, sample audio, environment configuration scripts (\texttt{requirements.txt}), etc. The \texttt{README.md} file provides an overview of the artifact and all detailed instructions for running the artifact. 

We do not include the original LibriSpeech, FMA, and VCTK datasets as they are too large in size and are publicly available for download. The scripts to preprocess the datasets are available in the repository. 
Due to space constraints, the artifact for our AudioMarkNet evaluation includes only the evaluation code. Specifically, we release the code in \texttt{amn\_code/}, which contains our modifications to AudioMarkNet's original \texttt{run\_evaluate.py} script for measuring watermark removal success rate against the AudioMarkNet detector, along with \texttt{prepare\_vctk.py} for preprocessing the VCTK dataset and \texttt{run\_trainSpeakerNet.py} for training the speaker verification model. Note that all the pretrained AudioMarkNet models and the full VCTK dataset are not included due to their large size. Instead, researchers can obtain the pretrained models and the same VCTK dataset required for reproducing the evaluation results from the AudioMarkNet released official codebase: \url{https://zenodo.org/records/14722182}, then run our evaluation code to obtain attack results.

\section*{Ethical Considerations} %

Advances in audio watermark removal are increasing the risks. For instance, they 
allow for the erasure of provenance from AI-generated speech, facilitating misinformation, impersonation, and fraud.
Still, researching watermark removal is ethically justified and important. By systematically studying the vulnerabilities of current watermarking schemes, we can identify weaknesses that might otherwise be exploited maliciously. Moreover, understanding the weaknesses of existing watermarking schemes enables the development of more robust watermarking techniques. Furthermore, publishing responsible findings encourages dataset and model owners to adopt stronger watermarking standards and promotes user awareness of the limitations of the existing protection mechanisms. Ultimately, such research contributes to a safer and more trustworthy digital ecosystem rather than undermining it. To mitigate misuse, our release is intended for research evaluation of audio watermark robustness. We present our method as a red-team tool for improving watermarking schemes, and we avoid presenting it for bypassing provenance in deployed services.

\section*{Generative AI Usage}
The authors used generative AI tools to assist in preparing this manuscript. Specifically, Grammarly and ChatGPT-5 were used to polish the writing, improve individual sentence clarity, and correct typographical and grammatical errors in the manuscript. Generative AI was not used to generate the core technical ideas, theoretical explanations, experimental claims, results, and main contributions of the manuscript.

The authors used Claude Code to assist with setting up the correct experimental environment, writing trivial test scripts, and anonymizing the repository during the double-blind review process. All AI-assisted code was reviewed by the authors, tested, and validated by checking that the scripts produced the expected experimental outputs.

The authors take full responsibility for the content of the manuscript, including all the technical claims, experimental results, citations, and code. All AI-assisted materials were manually reviewed, edited, and verified by the authors before including them in the manuscript.

\begin{figure*}[ht!]
    \centering
    \includegraphics[width=.63\linewidth]{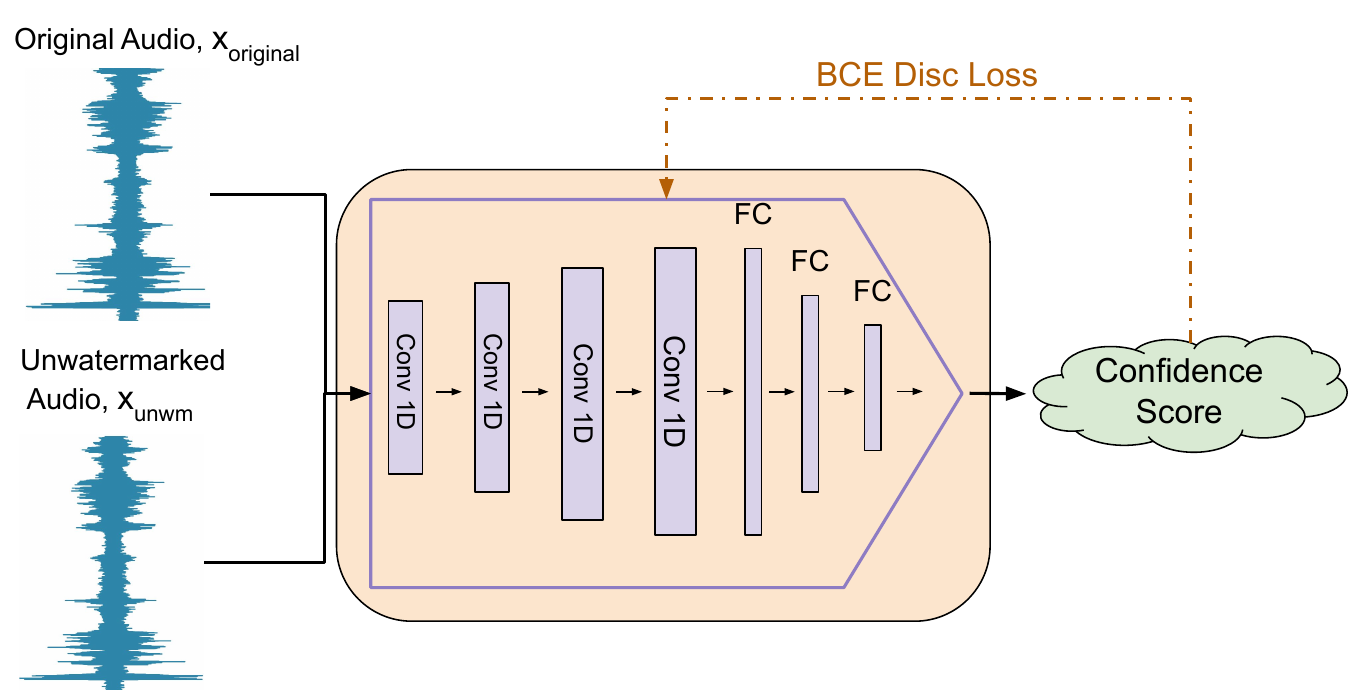}
    \caption{\tool's adversarial discriminator architecture.}
    \label{fig:disc}
\end{figure*}

\begin{algorithm}
\caption{\tool Training}
\label{alg:training}
\begin{algorithmic}[1]
\State \textbf{Input:} Dataset $\mathcal{D} = \{(x_i^{\text{original}}, x_i^{\text{wm}})\}_{i=1}^N$, epochs $E$,
\State \textbf{Output:} Trained autoencoder $G^*_\theta$, discriminator $D^*_\phi$
\State Initialize autoencoder $G_\theta$, discriminator $D_\phi$
\State Initialize optimizers $\text{opt}_G$, $\text{opt}_D$
\For{epoch $e = 1$ to $E$}
    \For{batch $(x_{\text{original}}, x_{\text{wm}})$ in $\mathcal{D}$}
        \State \Comment{Train discriminator}
        \State $x_{\text{unwm}} \gets G_\theta(x_{\text{wm}})$ \Comment{Generate unwatermarked audio}
        \State $\mathcal{L}_{\text{original}} \gets \text{BCE}(D_\phi(x_{\text{original}}), 1)$ 
        \State $\mathcal{L}_{\text{unwatermarked}} \gets \text{BCE}(D_\phi(\text{detach}(x_{\text{unwm}})), 0)$ 
        \State $\mathcal{L}_{\text{disc}} \gets \frac{1}{2}[\mathcal{L}_{\text{original}} + \mathcal{L}_{\text{unwatermarked}}]$
        \State Update $D_\phi$ with $\nabla_\phi \mathcal{L}_{\text{disc}}$

        \State \Comment{Train autoencoder}
        \State $\mathcal{L}_{\text{recon}} \gets |x_{\text{unwm}} - x_{\text{original}}| + 0.1(x_{\text{unwm}} - x_{\text{original}})^2$
        \State $\mathcal{L}_{\text{disentangle}} = \frac{1}{2}\left(1 + \text{cosine\_sim}(\Delta_{\text{proc}}, \Delta_{\text{wm}})\right)$
        \State $\mathcal{L}_{\text{local}} \gets \sum_{m=1}^{M} w_m \cdot r_m$
        \State 
        \State $\mathcal{L}_{\text{adv}} \gets \text{BCE}(D_\phi(x_{\text{unwm}}), 1)$ \Comment{Confuse discriminator}
        \State $\mathcal{L}_{\text{total}} \gets \alpha_r \mathcal{L}_{\text{recon}} + \alpha_l \mathcal{L}_{\text{local}} + \alpha_d \mathcal{L}_{\text{disentangle}} + \alpha_a \mathcal{L}_{\text{adv}}$
        \State Update $G_\theta$ with $\nabla_\theta \mathcal{L}_{\text{total}}$
    \EndFor
\EndFor
\State \Return $G^*_\theta$, $D^*_\phi$
\end{algorithmic}
\end{algorithm}

\begin{algorithm}
\caption{\tool Testing}
\label{alg:testing}
\begin{algorithmic}[1]
\State \textbf{Input:} Trained model $G^*_\theta$, watermarked audio $x_{\text{wm}}$, watermark detector $V$
\State \textbf{Output:} Processed audio $x_{\text{unwm}}$, attack success flag
\State $x_{\text{unwm}} \gets G^*_\theta(x_{\text{wm}})$ \Comment{Single forward pass}
\State $\text{conf}_{\text{before}} \gets V.\text{detect}(x_{\text{wm}})$ \Comment{Original detection confidence}
\State $\text{conf}_{\text{after}} \gets V.\text{detect}(x_{\text{unwm}})$ \Comment{Post-attack confidence}
\State $\text{success} \gets (\text{conf}_{\text{after}} < \tau)$ \Comment{$\tau$ is detection threshold}
\State \Return $x_{\text{unwm}}$, $\text{success}$
\end{algorithmic}
\end{algorithm}

\section{Detailed Discriminator Architecture of \tool} \label{app:disc}

Figure~\ref{fig:disc} demonstrates the detailed discriminator architecture discussed in Section~\ref{subsec:disc}.

\section{\tool{} Training and Testing Algorithms}\label{app:algo}

Algorithms~\ref{alg:training} and~\ref{alg:testing} describe the training and testing procedures of \tool{}, as discussed in Section~\ref{subsec:unified}. The watermark detector $V$ in Algorithm~\ref{alg:testing} is used only for evaluation and is not available to the attacker during watermark removal. The threshold $\tau$ represents an example of a high-level zero-bit detection rule; the detailed attack success criterion for each watermarking scheme is described in the Attack Success Rate paragraph in Section~\ref{sec:eval-transfer}.

\begin{table}[h!]
\centering
\caption{Average Attack Success Rate of Signal Distortion Attacks on AudioSeal watermarks.}
\label{tab:signal_editing}
\resizebox{0.7\columnwidth}{!}{
\begin{tabular}{lcc}
\toprule
\textbf{Signal Distortion} & \textbf{LibriSpeech} & \textbf{FMA} \\
\midrule
Bandpass Filter & 2\%  & 0\% \\
Lowpass Filter  & 10\% & 0\% \\
Highpass Filter & 6\%  & 0\% \\
Gaussian Noise  & 0\%  & 0\% \\
Pink Noise      & 2\%  & 0\% \\
Resampling       & 0\%  & 0\% \\
Echo Effect     & 0\%  & 0\% \\
Duck Effect     & 0\%  & 0\% \\
\bottomrule
\end{tabular}
}
\end{table}

\section{Performance of Baseline Signal Distortion Attacks}\label{app:signal}
As discussed in Section~\ref{sec:eval-transfer}, signal-level attacks are ineffective at removing AudioSeal watermarks, one of the representative watermarking candidates evaluated in our study (See Table~\ref{tab:signal_editing}).

\begin{figure*}[htbp]
    \centering

    \subfloat[Spectrogram of watermark signals added by AudioSeal.]
    {
        \includegraphics[width=0.31\textwidth]{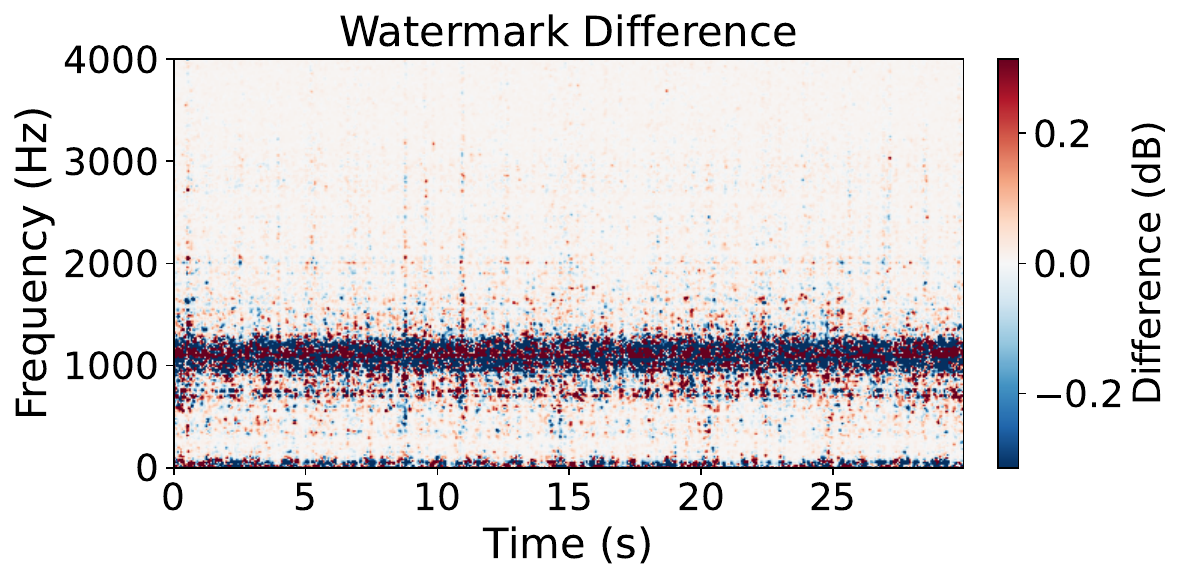}
        \label{fig:watermark_diff}
    }\hfill
    \subfloat[Spectrogram of \tool{} removal signals.]
    {
        \includegraphics[width=0.31\textwidth]{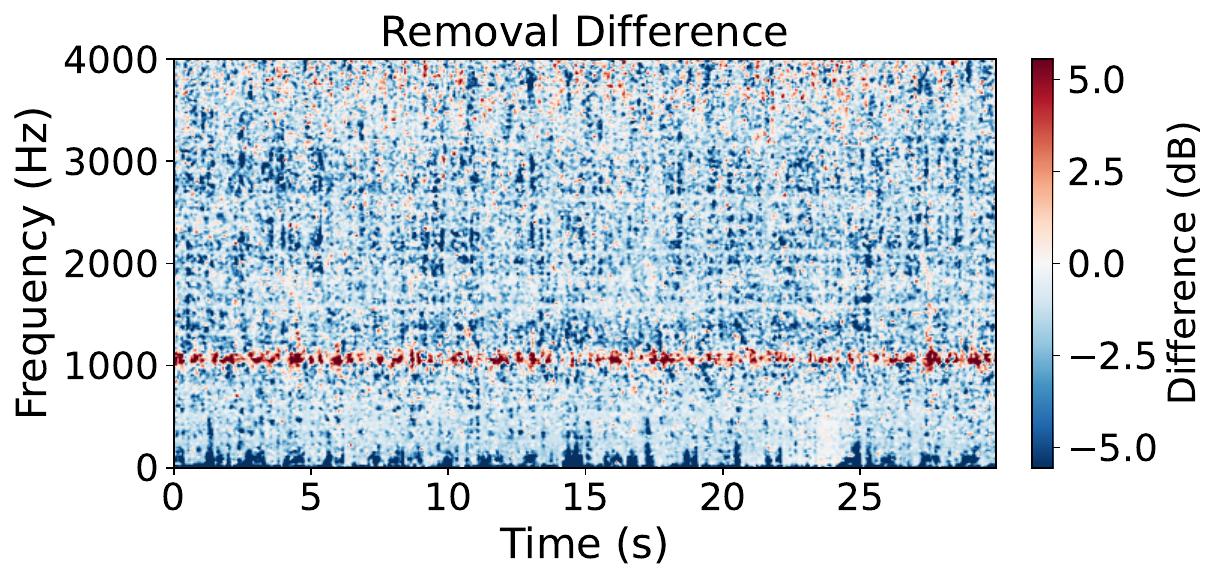}
        \label{fig:harmonicaudio_remove_diff_app}
    }\hfill
    \subfloat[Spectrogram of \audiomarkbench removal signals.]
    {
        \includegraphics[width=0.31\textwidth]{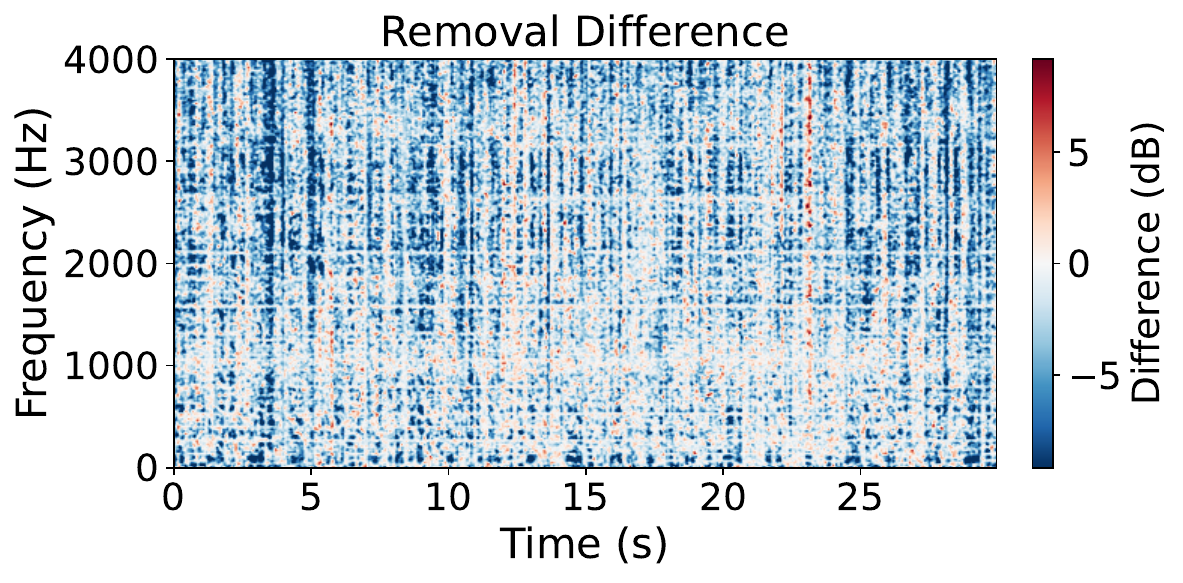}
        \label{fig:audiomarkbench_remove_diff}
    }

    \caption{Comparison of watermark signal spectrograms, \tool{} removal spectrograms, and \audiomarkbench removal spectrograms for AudioSeal-watermarked FMA audio. \tool is evaluated by transferring from the model trained on AudioSeal LibriSpeech samples.}
    \label{fig:three_difference_spectra}
\end{figure*}

\begin{figure*}[htbp]
    \centering

    \subfloat[Spectrogram of watermark signals added by WavMark.]
    {
        \includegraphics[width=0.31\textwidth]{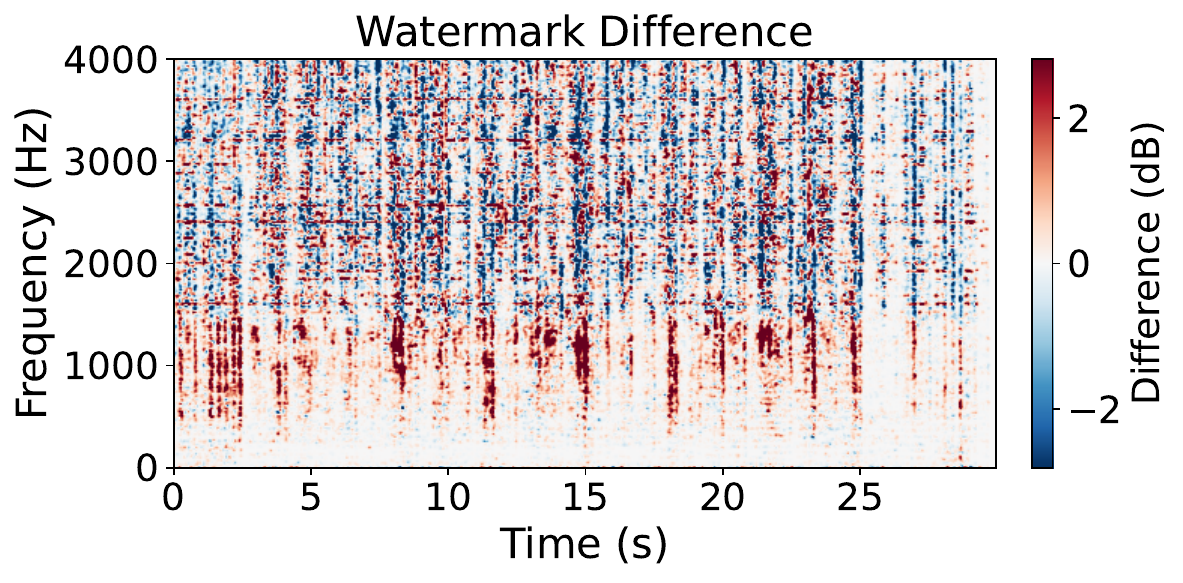}
        \label{fig:wavmark_watermark_diff}
    }\hfill
    \subfloat[Spectrogram of \tool{} removal signals.]
    {
        \includegraphics[width=0.31\textwidth]{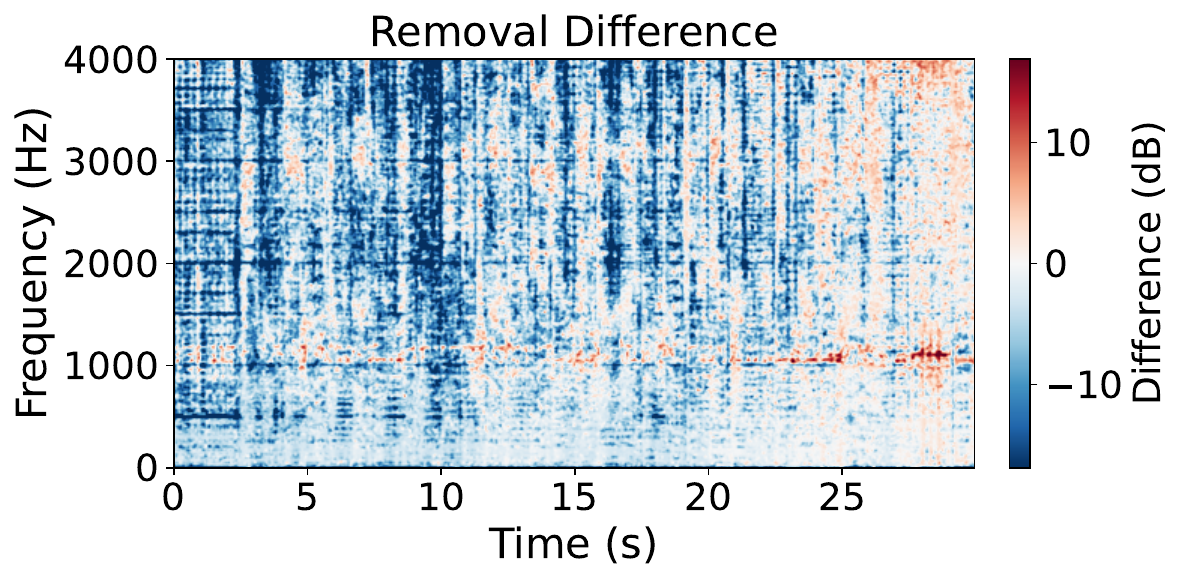}
        \label{fig:wavmark_harmonicaudio_remove_diff_app}
    }\hfill
    \subfloat[Spectrogram of \audiomarkbench removal signals.]
    {
        \includegraphics[width=0.31\textwidth]{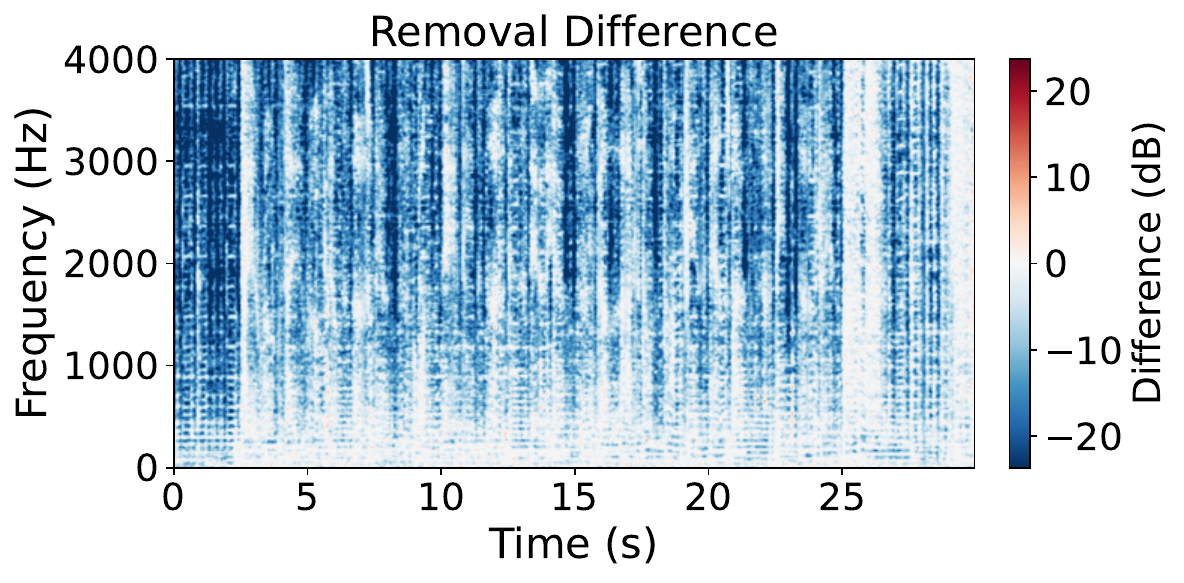}
        \label{fig:wavmark_audiomarkbench_remove_diff}
    }

    \caption{Comparison of watermark signal spectrograms, \tool{} removal spectrograms, and \audiomarkbench removal spectrograms for WavMark-watermarked FMA audio. \tool is evaluated by transferring from the model trained on AudioSeal LibriSpeech samples.}
    \label{fig:three_difference_spectra_wavmark}
\end{figure*}

\section{Comparison Between Watermarked and Watermark-Removed Spectrograms}
\label{appendix:watermark_remove_diff}
The results in this section complement the spectrogram analyses discussed in Section~\ref{subsec:transfer}. Instead of plotting the full audio signals, we focus on the watermark residuals before and after removal by \tool and \audiomarkbench. Figure~\ref{fig:three_difference_spectra} corresponds to Figure~\ref{fig:three_spectrograms}, and Figure~\ref{fig:three_difference_spectra_wavmark} corresponds to Figure~\ref{fig:WavMark_three_spectrograms_WavMark}.

\end{document}